\documentclass[twocolumn,aps,prc,showpacs,superscriptaddress]{revtex4}
\usepackage{psfig}

\newcommand{ \be	}{\begin{equation}}
\newcommand{ \ee 	}{\end{equation}}
\newcommand{ \bea 	}{\begin{eqnarray}}
\newcommand{ \eea	}{\end{eqnarray}}
\newcommand{ \ba	}{\begin{array}}
\newcommand{ \ea 	}{\end{array}}
\newcommand{ \bw	}{\begin{widetext}}
\newcommand{ \ew	}{\end{widetext}}

\newcommand{ \dphi	}{\Delta\phi}
\newcommand{ \phitrig	}{\phi_{\rm t}}
\newcommand{ \vtrig	}{v_2^{\rm (t)}}
\newcommand{ \vvtrig	}{v_4^{\rm (t)}}
\newcommand{ \vone	}{v_2^{\rm (1)}}
\newcommand{ \vvone	}{v_4^{\rm (1)}}
\newcommand{ \vtwo	}{v_2^{\rm (2)}}
\newcommand{ \vvtwo	}{v_4^{\rm (2)}}

\newcommand{ \pt	}{p_{\perp}}
\newcommand{ \mean	}[1]{\langle {#1} \rangle}
\newcommand{ \Jhat	}{\hat{J}}
\newcommand{ \rhohat	}{\hat{\rho}}
\newcommand{ \Btrue	}{B_{\rm true}}

\begin{document}

%%%%%%%%%%%%%%%%%%%%%%%%%%%%%%%%%%%%%%%%%%%%%%%%%%%%%%%%%%%%%%%%%%%%%%

\title{Cumulant versus jet-like three-particle correlations}

\author{Jason G. Ulery}
\affiliation{Department of Physics, Purdue University, 525 Northwestern Avenue, West Lafayette, Indiana 47907, USA}
\affiliation{Institut f\"{u}r Kernphysik, Max-von-Laue-Strasse 1, D-60438 Frankfurt am Main, Germany}
\author{Fuqiang Wang}
\affiliation{Department of Physics, Purdue University, 525 Northwestern Avenue, West Lafayette, Indiana 47907, USA}

\begin{abstract}
Two-particle jet-like azimuthal correlations have revealed intriguing modifications to the away-side of high $\pt$ trigger particles in relativistic heavy-ion collisions. Three-particle jet-like azimuthal correlation and three-particle azimuthal cumulant have been analyzed in experiments in attempt to distinguish conical emission of jet-correlated particles from other physics mechanisms. We investigate the difference between three-particle jet-like correlation and three-particle cumulant in azimuth. We show, under the circumstance where the away-side two-particle correlation is relatively flat in azimuth and similar in magnitude to the azimuthal average of the two-particle correlation signal, that the three-particle cumulant cannot distinguish conical emission from other physics mechanisms. The three-particle jet-like correlation, on the other hand, retains its discrimination power.
\end{abstract}


\maketitle

%%%%%%%%%%%%%%%%%%%%%%%%%%%%%%%%%%%%%%%%%%%%%%%%%%%%%%%%%%%%%%%%%%%%%%
\section{Introduction}

Two-particle jet-like azimuthal correlations have revealed significant modification in central heavy-ion collisions at RHIC~\cite{b2b,jetspec,Wang,Phenix,StarQM05,PhenixQM05}. The away-side particles associated with and opposite to a high transverse momentum ($\pt$) trigger particle are found to be broadly distributed about $\dphi=\pi$ in azimuth from the trigger particle, in contrast to observations from $pp$ and $d$+Au collisions. The shape of the broad away-side distribution varies with the associated particle $\pt$. For $1<\pt<2$~GeV/$c$, for instance, the away-side distribution may even be double-humped with a dip at $\dphi=\pi$~\cite{jetspec,Wang,Phenix,StarQM05,PhenixQM05}. The away-side associated particles are also found to be not much harder than the bulk medium particles~\cite{jetspec,Wang,Phenix,StarQM05}. The particles at $\dphi=\pi$ are found to be softer than those in the angular regions where the humps appear~\cite{Wang,Phenix,StarQM05}, again in contrast to observations in $pp$ and $d$+Au collisions or jet fragmentation in vacuum.

Several physical scenarios are possible to explain the observations. One is that jets may be deflected by radial transverse flow of the bulk medium or by the larger survival probability of jet particles moving outwards than inwards due to energy loss~\cite{Hwa}. Such a scenario would have jet particles narrowly clustered in individual events but the cluster is randomly distributed about $\dphi=\pi$ over many events. The second is large angle gluon radiation~\cite{Vitev}. This scenario would have qualitatively similar structure as for deflected jets. The third is conical flow from sound shock-waves generated by large energy deposition of high momentum partons and its strong pressure disturbance in the medium~\cite{MachCone}. Such shock-waves result in a distinctive Mach-cone type structure where particles are preferentially emitted at a Mach angle determined by the speed of sound in the medium, independent of the particle $\pt$. If Mach-cone type conical flow is indeed responsible for the observation, then the extraction of the speed of sound may be possible, thereby the equation of state of the created medium. The fourth is $\check{\rm C}$erenkov gluon radiation generated by interactions of fast moving particles with the medium~\cite{Cherenkov}. Such a scenario would have a similar structure as for the Mach-cone conical flow, but the $\check{\rm C}$erenkov angle will likely depend on the associated particle $\pt$~\cite{Cherenkov}.

Two-particle correlation cannot distinguish these scenarios because they give qualitatively the same two-particle correlation. Such ambiguity is lifted in three-particle correlation and its $\pt$ dependence. If the broad two-particle correlation is due to deflected jets or large angle gluon radiation, the two associated particles will be narrowly clustered in angle but the cluster will swing over a wide range in azimuth on the away side. If the Mach-cone or $\check{\rm C}$erenkov radiation is responsible for the broad two-particle correlation, then the two associated particles will have equal probability being opposite away from $\dphi=\pi$ as being clustered together. A three-particle correlation signal with opposite azimuthal angles from $\dphi=\pi$ for the two associated particles is, therefore, a distinctive signature of Mach-cone conical flow or $\check{\rm C}$erenkov radiation. The $\pt$ dependence of the cone angle may further discriminate between the two scenarios of Mach-cone conical flow and $\check{\rm C}$erenkov gluon radiation~\cite{Cherenkov}.

Three-particle jet-like azimuthal correlations have been studied in STAR~\cite{Star_3part}. The results show evidence of conical emission. The emission angle is independent of associated particle $\pt$ and is consistent with  Mach-cone shock waves~\cite{MachCone}, but not with simple $\check{\rm C}$erenkov gluon radiation~\cite{Cherenkov}. The analysis followed the jet-correlation method commonly used in two-particle azimuthal correlation studies at RHIC~\cite{b2b,jetspec,Wang,Phenix,StarQM05,PhenixQM05}, but extended to three particles. Hereon we will refer to this method as the {\em jet-correlation method}. Recently another analysis method, the three-particle cumulant method with azimuthal angle defined in the laboratory frame~\cite{cumulant}, has been proposed. The method follows the mathematically well-defined cumulant concept. Hereon we will refer to this method as the {\em lab-frame cumulant method}. The two methods give different results and may confuse the general reader. In this paper we shall compare the two methods and discuss their differences in detail. We first give brief descriptions of the two methods. We then compare the two methods and discuss their differences, using a simple analytical model for jets. Finally we draw our conclusions.

%%%%%%%%%%%%%%%%%%%%%%%%%%%%%%%%%%%%%%%%%%%%%%%%%%%%%%%%%%%%%%%%%%%%%%
\section{Descriptions of the Two Analysis Methods}

The objectives of the two analysis methods, the jet-correlation method and the cumulant method, are both to study jet structures. Due to the large particle multiplicity in relativistic heavy-ion collisions, event-by-event reconstruction of jets is impossible; one often resorts to two- and three-particle azimuthal correlations of charged hadrons with high $\pt$ trigger particles that have a relatively large probability to originate from dijets. The obtained correlation functions, thus, yield information on dijets. In this section, we briefly describe the two analysis methods for three-particle azimuthal correlation studies.

%%%%%%%%%%%%%%%%%%%%%%%%%%%%%%%%%%%%%%%%%%%%%%%%%%%%%%%%%%%%%%%%%%%%%%
\subsection{The jet-correlation method}

The three-particle jet-correlation method is described in~\cite{method,Star_3part}. The method is extended from the commonly used two-particle jet-correlation method to three particles. Combinatorial backgrounds are obtained from event-mixing technique; they include background from three ``random'' particles as well as background from a correlated trigger-associated particle pair with a ``random'' associated particle. The ``random'' particles we refer to here (and hereafter without the quotes) may include other correlations that are not related to the trigger particle, such as those due to anisotropic flow.

The jet-like correlation method has the jet model in mind. The difficulty is that the underlying background is unknown a priori. One has to make ad hoc working assumptions about the background level. The common assumptions made in data analysis are ZYA1~\cite{jetspec,Wang,Phenix,StarQM05,method,Star_3part} and ZYAM~\cite{Phenix,PhenixQM05}. The correlation measured at RHIC is the lowest around $\dphi=\pm 1$. The STAR experiment makes the assumption that the jet signal is zero within the fixed range of $0.8<|\dphi|<1.2$ (ZYA1)~\cite{jetspec}, while the PHENIX experiment uses the so-called zero-yield-at-minimum (ZYAM) method in which the $\dphi$ region where the signal minimum resides is determined by the data itself~\cite{Phenix}. 

The two-particle jet-like correlation is 
\be\Jhat_2(\dphi)=J_2(\dphi)-B_2(\dphi),\label{eq5}\ee
where $\dphi=\phi-\phitrig$ is the azimuthal angle difference between associated and trigger particles. $J_2(\dphi)$ is the two-particle raw correlation function between the trigger and associated particles, and $B_2(\dphi)$ is the combinatorial background normalized by the aforementioned normalization schemes. The three-particle jet-like correlation is
\be
\ba{ll}
\Jhat_3(\dphi_1,\dphi_2)=&J_3(\dphi_1,\dphi_2)-B_3(\dphi_1,\dphi_2)\\
&-\Jhat_2(\dphi_1)B_2(\dphi_2)-\Jhat_2(\dphi_2)B_2(\dphi_1),
\ea
\label{eq6a}
\ee
or alternatively
\be
\ba{ll}
\Jhat_3(\dphi_1,\dphi_2)=&J_3(\dphi_1,\dphi_2)-B_3(\dphi_1,\dphi_2)\\
&-J_2(\dphi_1)B_2(\dphi_2)-J_2(\dphi_2)B_2(\dphi_1)\\
&+2B_2(\dphi_1)B_2(\dphi_2),
\ea
\label{eq6b}
\ee
where $\dphi_i=\phi_i-\phitrig (i=1,2)$ are the azimuthal angles of the associated particles relative to that of the trigger particle. $J_3(\dphi_1,\dphi_2)$ is the three-particle raw correlation function. $B_3(\dphi_1,\dphi_2)$ is the combinatorial background of three-particle correlation between two random associated particles with a random trigger particle. The third and fourth terms in the r.h.s. and Eqs.~(\ref{eq6a}) and (\ref{eq6b}) are the other background, the combinatorial background of a correlated trigger-associated pair with a random associated particle. It is given by the product of the two-particle jet-correlation signal $\Jhat_2$ with the underlying background $B_2$.

In data analysis~\cite{jetspec,Wang,Phenix,StarQM05,PhenixQM05,Star_3part}, the dihadron background $B_2(\dphi)$ is often obtained from the mixed-event technique, mixing a trigger particle from one event with an associated particle from another event, and then scaling the result by a normalization factor, $a$, using the ZYA1 or ZYAM scheme. The $B_3(\dphi_1,\dphi_2)$ background is obtained from the mixed-event technique using three different events, with a proper normalization scale, $a^2b$, where $b$ quantifies the difference in the degrees of deviation from Poisson statistics in the background associated particle multiplicity and that from mixed-events~\cite{Star_3part}. In this paper, for simplicity we shall constrain ourselves to $b=1$.

The advantage of the jet-correlation method is that, once the assumption about the background is made and the level of background is determined, the resultant three-particle jet-like correlation signal is easy to interpret and can be used to discriminate different physics scenarios. The disadvantage is of course the difficulty of the analysis.

%%%%%%%%%%%%%%%%%%%%%%%%%%%%%%%%%%%%%%%%%%%%%%%%%%%%%%%%%%%%%%%%%%%%%%
\subsection{The lab-frame cumulant method}

The lab-frame three-particle cumulant method is described in detail in~\cite{cumulant}. Given a high $\pt$ trigger particle to preferentially select a dijet, the two-particle cumulant is defined as
\be\rhohat_2(\phitrig,\phi)=\rho_2(\phitrig,\phi)-\rho_1(\phitrig)\rho_1(\phi),\label{eq1}\ee
where $\phitrig$ and $\phi$ are azimuthal angles of the trigger and associated particles in the {\em laboratory frame}, respectively. The three-particle cumulant is defined as
\bea
\rhohat_3(\phitrig,\phi_1,\phi_2)&=&\rho_3(\phitrig,\phi_1,\phi_2)-\rho_2(\phi_1,\phi_2)\rho_1(\phitrig)\nonumber\\
&&-\rho_2(\phitrig,\phi_1)\rho_1(\phi_2)-\rho_2(\phitrig,\phi_2)\rho_1(\phi_1)\nonumber\\
&&+2\rho_1(\phitrig)\rho_1(\phi_1)\rho_1(\phi_2),\label{eq2}
\eea
where $\phi_i$ is the azimuthal angle of the $i$th associated particle ($i=1,2$) in the laboratory frame. In Eq.~(\ref{eq2}), $\rho_1(\phi)=dN/d\phi$ is the average single particle density, $\rho_2(\phi_1,\phi_2)=d^2N/d\phi_1d\phi_2$ is the average two-particle density, and $\rho_3(\phitrig,\phi_1,\phi_2)=d^3N/d\phitrig d\phi_1d\phi_2$ is the average three-particle density. 

Cumulants are normally computed on a per event basis. Since our goal here is to identify jet-correlation structure, we normalize the cumulants by the number of trigger particles. The two normalizations differ simply by a constant factor which is the average number of trigger particles per event. 

It is worth to note here that $\rho_1$, $\rho_2$, and $\rho_3$ are all event-wise average quantities. They depend on what event sample is used in the analysis. Consider for example two event samples within a given centrality bin: minimum bias event sample and only events containing a trigger particle (triggered events). The quantities related to a trigger particle, $\rho_3(\phitrig,\phi_1,\phi_2)$, $\rho_2(\phitrig,\phi_1)$, $\rho_2(\phitrig,\phi_2)$, and $\rho_1(\phitrig)$ are all zero in the nontriggered events, and thus remain the same between the two event samples if normalized per trigger particle (or differ by a constant if normalized per event). The other quantities, $\rho_2(\phi_1,\phi_2)$, $\rho_1(\phi_1)$, and $\rho_1(\phi_2)$ are generally different between triggered events and nontriggered events due to trigger bias. Thus the cumulants analyzed using two different event samples differ, and this is beyond the simple normalization factor between per trigger and per event normalizations mentioned above. In other words, the cumulant analyzed using only the triggered events yields $\rhohat_3$(triggered events), the cumulant analyzed using only the nontriggered events would yield zero by definition, and the cumulant analyzed using both the triggered and nontriggered events together yields $\rhohat_3$(all events), and $\rhohat_3$(triggered events) $\neq$ $\rhohat_3$(all events). Moreover, $\rhohat_3$(all events) depends on the relative numbers of nontriggered events (for which $\rhohat_3\equiv0$) and triggered events. In this paper, we shall constrain ourselves to using triggered events only.

The per trigger normalized two-particle cumulant is
\be\rhohat_2(\dphi)=\rho_2(\dphi)-\rho_1(\dphi),\label{eq3}\ee
and the three-particle cumulant is
\bea
\rhohat_3(\dphi_1,\dphi_2)&=&\rho_3(\dphi_1,\dphi_2)-\rho_2(\dphi_1,\dphi_2)\nonumber\\
&&-\rho_2(\dphi_1)\rho_1(\dphi_2)-\rho_2(\dphi_2)\rho_1(\dphi_1)\nonumber\\
&&+2\rho_1(\dphi_1)\rho_1(\dphi_2),\label{eq4a}
\eea
or alternatively
\be\ba{ll}
\rhohat_3(\dphi_1,\dphi_2)=&\rho_3(\dphi_1,\dphi_2)-\rho_2(\dphi_1,\dphi_2)\\
&-\rhohat_2(\dphi_1)\rho_1(\dphi_2)-\rhohat_2(\dphi_2)\rho_1(\dphi_1).\label{eq4b}
\ea\ee
In Eqs.~(\ref{eq3}), (\ref{eq4a}) and (\ref{eq4b}), $\rho_2(\dphi)=dN/d\dphi$ is the two-particle raw correlation function (or, equivalently, associated single-particle density per trigger particle in $\dphi=\phi-\phitrig$), and $\rho_3(\dphi_1,\dphi_2)=d^2N/d\dphi_1d\dphi_2$ is the three-particle raw correlation function (or, equivalently, associated pair-density per trigger particle in $\dphi_i=\phi_i-\phitrig$). In addition, $\rho_2(\dphi_1,\dphi_2)\equiv\rho_2(\phi_1,\phi_2)=d^2N/d\phi_1d\phi_2$ is the two-particle density but expressed in the azimuth differences of the particles from a random trigger particle.

Because the azimuthal angles are defined in the laboratory frame, the single particle density $\rho_1(\phi)$ is constant and we shall just use $\rho_1$:
\be
\rho_1(\phi)=\rho_1(\dphi)=\rho_1={\rm const.}
\ee
The reason that $\rho_1(\dphi)$ is also constant is because the product $\rho_1(\phitrig)\rho_1(\phi)$ is taken after the $\rho_1$'s have been already averaged over the event sample.

Cumulants are well defined mathematically. The data analysis of cumulants is straightforward. One calculates the cumulants for each event and accumulates them over a event sample; the cumulants can be binned in fixed-multiplicity bins. The shortcoming, as we will see, is the difficulty in the interpretation of the obtained results: the cumulants are mathematically well defined and thus do not depend on the underlying physics of the events; their interpretations, therefore, have to depend on the model for the underlying physics. 

%%%%%%%%%%%%%%%%%%%%%%%%%%%%%%%%%%%%%%%%%%%%%%%%%%%%%%%%%%%%%%%%%%%%%%
\subsection{The reaction-plane-frame cumulant method}

Reaction plane direction in nucleus-nucleus collisions is random. The natural azimuthal angle to use is that defined relative to reaction plane, $\phi-\psi$. The formulism in the previous section all applies with a simple change:
\be
\phi\rightarrow\phi-\psi
\ee
and no change to $\dphi$. However, $\rho_1$ and $\rho_2$ are not constant anymore,
\bea
\rho_1(\phi)&\neq&{\rm const.},\nonumber\\
\rho_1(\dphi)&\neq&{\rm const.},
\eea
but rather are functions of $\phi-\psi$ and $\dphi$, respectively.

%%%%%%%%%%%%%%%%%%%%%%%%%%%%%%%%%%%%%%%%%%%%%%%%%%%%%%%%%%%%%%%%%%%%%%
\section{Comparisons between the Two Methods}

The lab-frame cumulant and the reaction-plane-frame cumulant are clearly different. This is because the cumulant is a mathematical construction, and an event sample expressed in lab-frame azimuthal angle and the same event sample expressed in azimuthal angle relative to reaction plane are different mathematical identities. However, the event sample contains obviously the same physics, whether expressed in the laboratory frame or in the reaction plane frame. In other words, the same physics yields two different cumulant results. Therefore, extreme care should be taken in the physics interpretations of cumulant results. 

The raw two- and three-particle correlation functions in the cumulant methods and in the jet-correlation method are identical, namely,
\be\rho_2(\dphi)\equiv J_2(\dphi),\label{eq7a}\ee
\be\rho_3(\dphi_1,\dphi_2)\equiv J_3(\dphi_1,\dphi_2).\label{eq7b}\ee
Comparison of Eqs.(\ref{eq6b}) and (\ref{eq4a}) reveals that the reaction-plane-frame cumulant and the jet-correlation method are identical if the background $B_2(\dphi)$ and $B_3(\dphi_1,\dphi_2)$ are taken from mixed-events without normalization scalings (i.e., $a=b=1$). This is because the same anisotropic flow is included in both methods. We shall therefore only focus on the comparison between the lab-frame cumulant method and the jet-correlation method.

We shall first compare the two methods in a simple case, in which the background particle distribution is uniform in azimuth (i.e., no anisotropic flow). Such a comparison is enlightening as the difference between the two methods is straightforward and can be easily identified. We then include anisotropic flow in the comparison. We note that there could be  intrinsic correlations (beyond those due to anisotropic flow) between the two associated particles in the background. Those correlations are not important for our discussion here, so for simplicity we assume no other intrinsic correlations between the two associated particles except those from anisotropic flow.

%%%%%%%%%%%%%%%%%%%%%%%%%%%%%%%%%%%%%%%%%%%%%%%%%%%%%%%%%%%%%%%%%%%%%%
\subsection{Jet-correlation with uniform background}

Since we consider no intrinsic correlation between the background particles and no anisotropic flow, we have 
\be\rho_2(\dphi_1,\dphi_2)\equiv\rho_2(\phi_1,\phi_2)=\rho_1^2.\label{eq8}\ee
Here we have assumed that the number of pairs equals to the square of the number of particles (Poisson statistics). Equations~(\ref{eq4a}) and (\ref{eq4b}) become
\bea
\rhohat_3(\dphi_1,\dphi_2)&=&\rho_3(\dphi_1,\dphi_2)-\rho_2(\dphi_1)\rho_1\nonumber\\
&&-\rho_2(\dphi_2)\rho_1+\rho_1^2,\label{eq9a}
\eea
and
\bea
\rhohat_3(\dphi_1,\dphi_2)&=&\rho_3(\dphi_1,\dphi_2)-\rhohat_2(\dphi_1)\rho_1\nonumber\\
&&-\rhohat_2(\dphi_2)\rho_1-\rho_1^2.\label{eq9b}
\eea
	
Since no anisotropic flow is considered, the background distribution in jet-correlation method is uniform,
\be B_2(\dphi)=B_1,\label{eq10a}\ee
where $B_1$ is the average background single-particle density. Since we consider no intrinsic correlation between the two background particles, we have
\be B_3(\dphi_1,\dphi_2)=B_1^2.\label{eq10b}\ee
Thus Eqs.~(\ref{eq6a}) and (\ref{eq6b}) become
\bea
\Jhat_3(\dphi_1,\dphi_2)&=&J_3(\dphi_1,\dphi_2)-\Jhat_2(\dphi_1)B_1\nonumber\\
&&-\Jhat_2(\dphi_2)B_1-B_1^2,\label{eq11a}
\eea
and
\bea
\Jhat_3(\dphi_1,\dphi_2)&=&J_3(\dphi_1,\dphi_2)-J_2(\dphi_1)B_1\nonumber\\
&&-J_2(\dphi_2)B_1+B_1^2.\label{eq11b}
\eea

The single particle densities in the two methods are different: $\rho_1$ is larger than $B_1$, and 
\be\rho_1-B_1=\mean{\Jhat_2},\label{eq12}\ee
where $\mean{\Jhat_2}$ is the average jet-correlated associated particle multiplicity density in azimuth. As we shall see, this difference is the essential piece that makes the results from the two methods differ. (Note that each event in our event sample contains a trigger particle. If events with no trigger particles are mixed into the event sample, then the difference $\rho_1-B_1$ becomes rather arbitrary depending on the relative mixer and the specifics of trigger bias. See detailed discussion in the previous section.)

Using Eqs.~(\ref{eq5}), (\ref{eq7a}), (\ref{eq7b}), (\ref{eq10a}), and (\ref{eq12}), the difference between the three-particle cumulant, Eq.~(\ref{eq9a}), and the three-particle jet-correlation, Eq.~(\ref{eq11b}), is given by 
\bea
\Delta
&=&\rhohat_3(\dphi_1,\dphi_2)-\Jhat_3(\dphi_1,\dphi_2)\nonumber\\
&=&-\mean{\Jhat_2}\left[\Jhat_2(\dphi_1)+\Jhat_2(\dphi_2)-\mean{\Jhat_2}\right].\label{eq13}
\eea

To give further insights, we consider the simple case in which the three-particle jet-correlation function is simply the product of the two two-particle jet-correlation functions (i.e., the three-particle jet-correlation function is factorized):
\be\Jhat_3(\dphi_1,\dphi_2)=\Jhat_2(\dphi_1)\Jhat_2(\dphi_2).\label{eq:Jhat3factorize}\ee
In other words, the two associated particles are not intrinsically correlated, but are correlated due to their individual correlations to the same trigger particle. This corresponds to the following extreme jet fragmentation scenario: the trigger particle direction is the jet axis, and fragmentations into individual hadrons are identical and independent of each other. Then, using Eq.~(\ref{eq13}), the three-particle cumulant is simply
\be\rhohat_3(\dphi_1,\dphi_2)=\left[\Jhat_2(\dphi_1)-\mean{\Jhat_2}\right]\left[\Jhat_2(\dphi_2)-\mean{\Jhat_2}\right].\label{eq15}\ee
This should come as no surprise; the three-particle cumulant can also be factorized into two two-particle cumulants that are given by Eq.~(\ref{eq3}). The difference between the two methods is in the background level. The jet-correlation method puts the background at ZYA1 or ZYAM, and the cumulant method, by definition, effectively takes the average of the raw correlation signal as background. In fact, if the jet-correlation method also takes the average as the background, i.e., the two-particle jet-correlation signal is now $\Jhat_2(\dphi)-\mean{\Jhat_2}$ instead of $\Jhat_2(\dphi)$, then it will yield the identical result as that from the cumulant method in Eq.~(\ref{eq15}).

\begin{figure*}[hbt]
\centerline{
\psfig{file=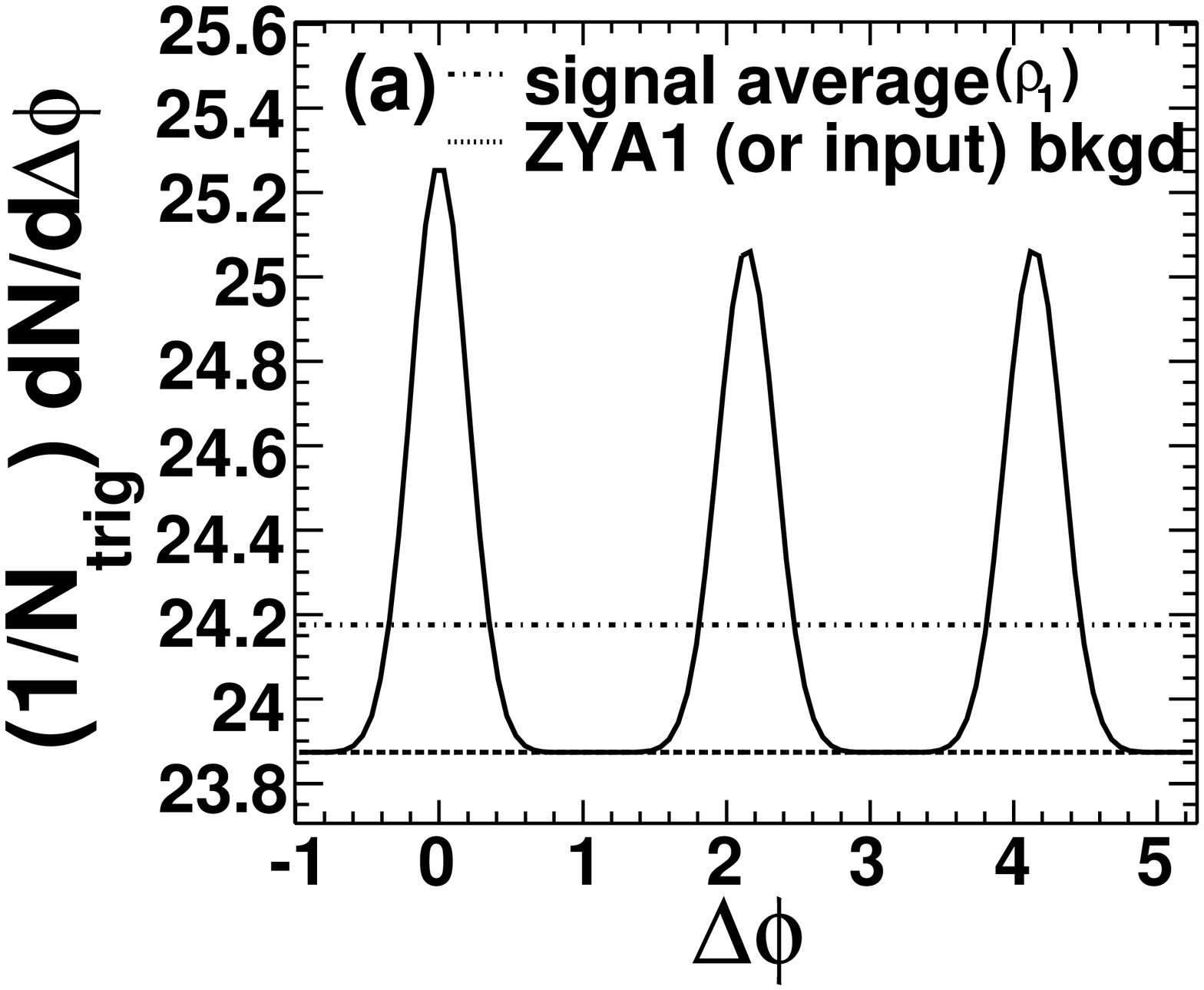,width=0.25\textwidth}
\psfig{file=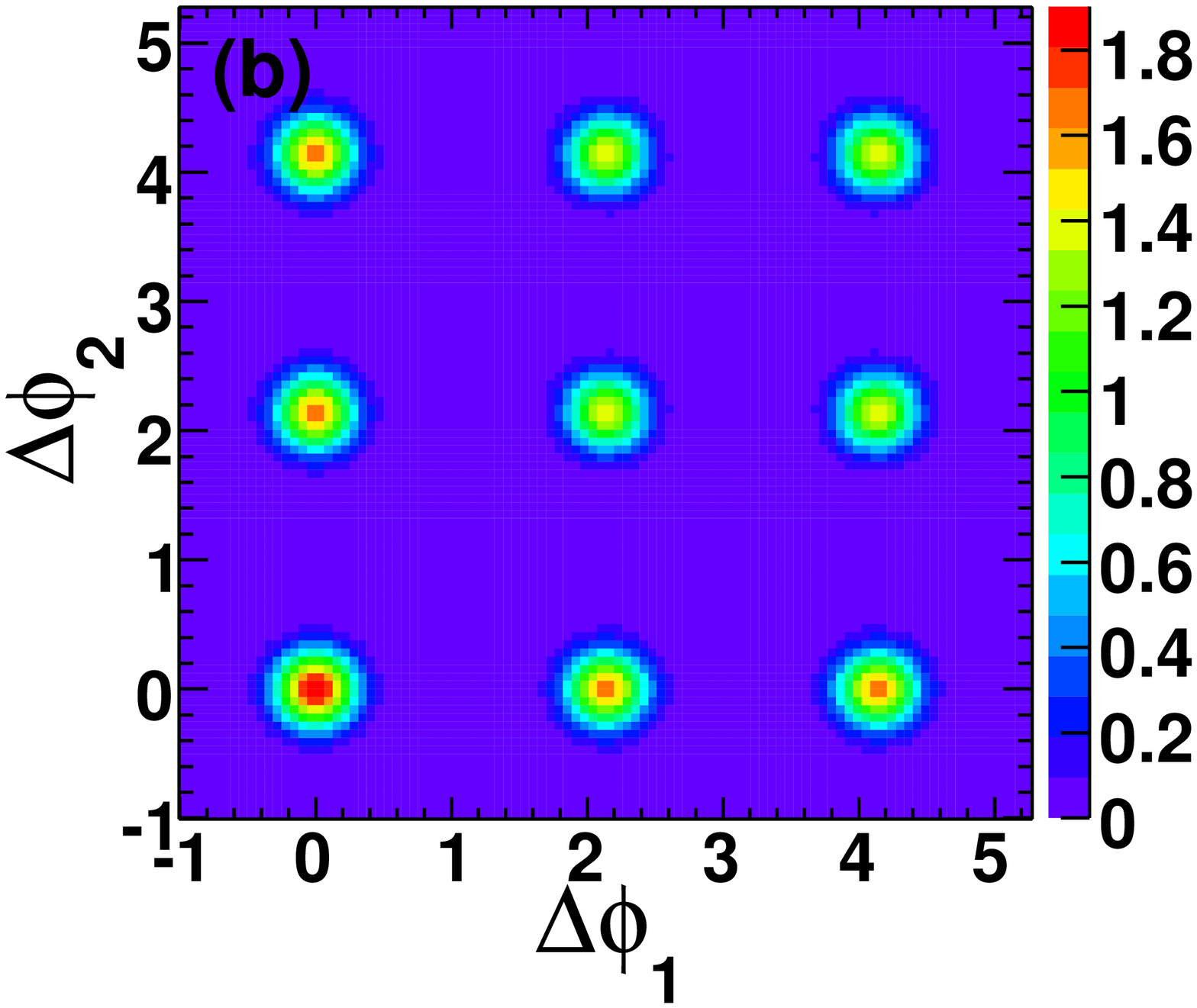,width=0.25\textwidth}
\psfig{file=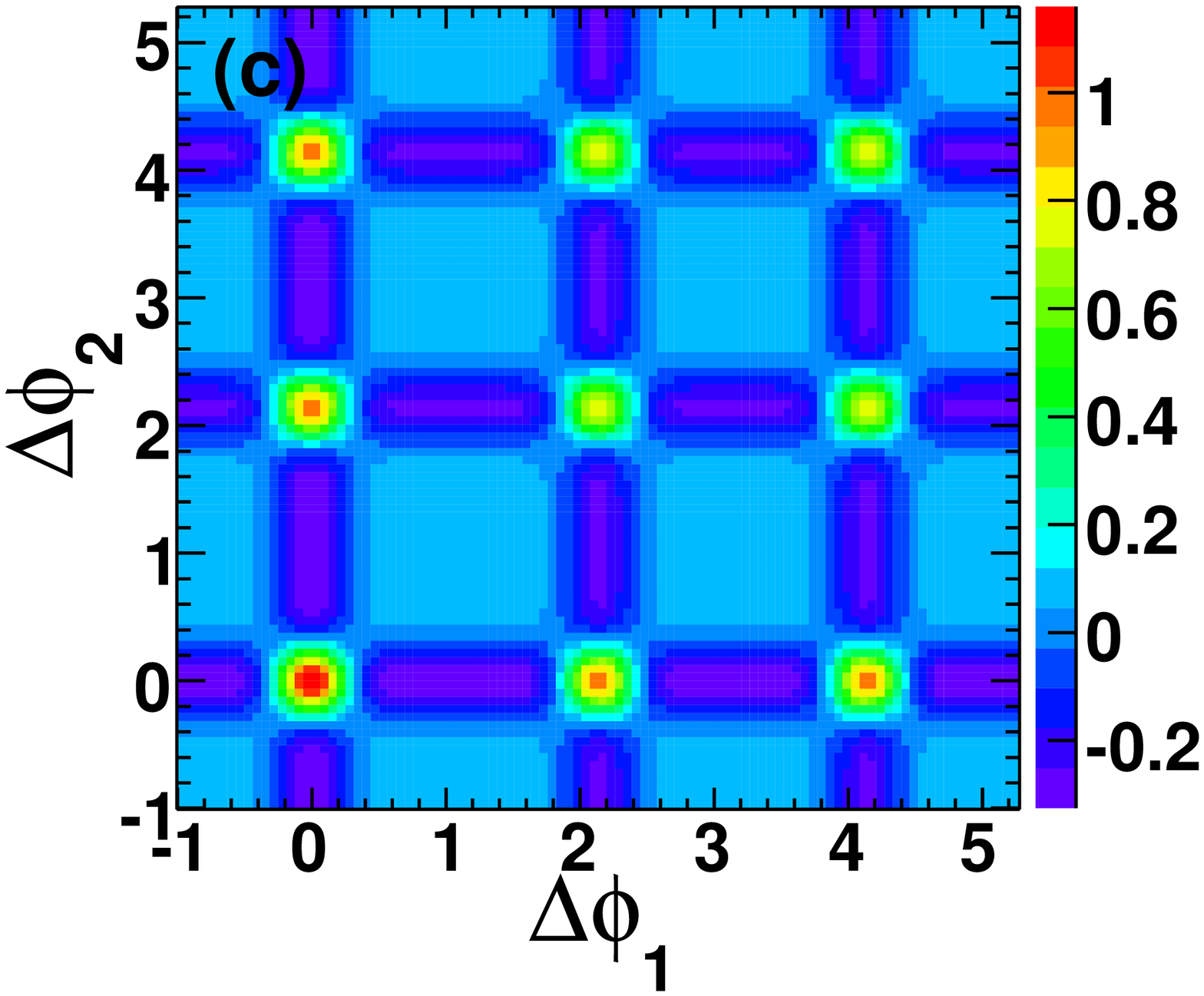,width=0.25\textwidth}
\psfig{file=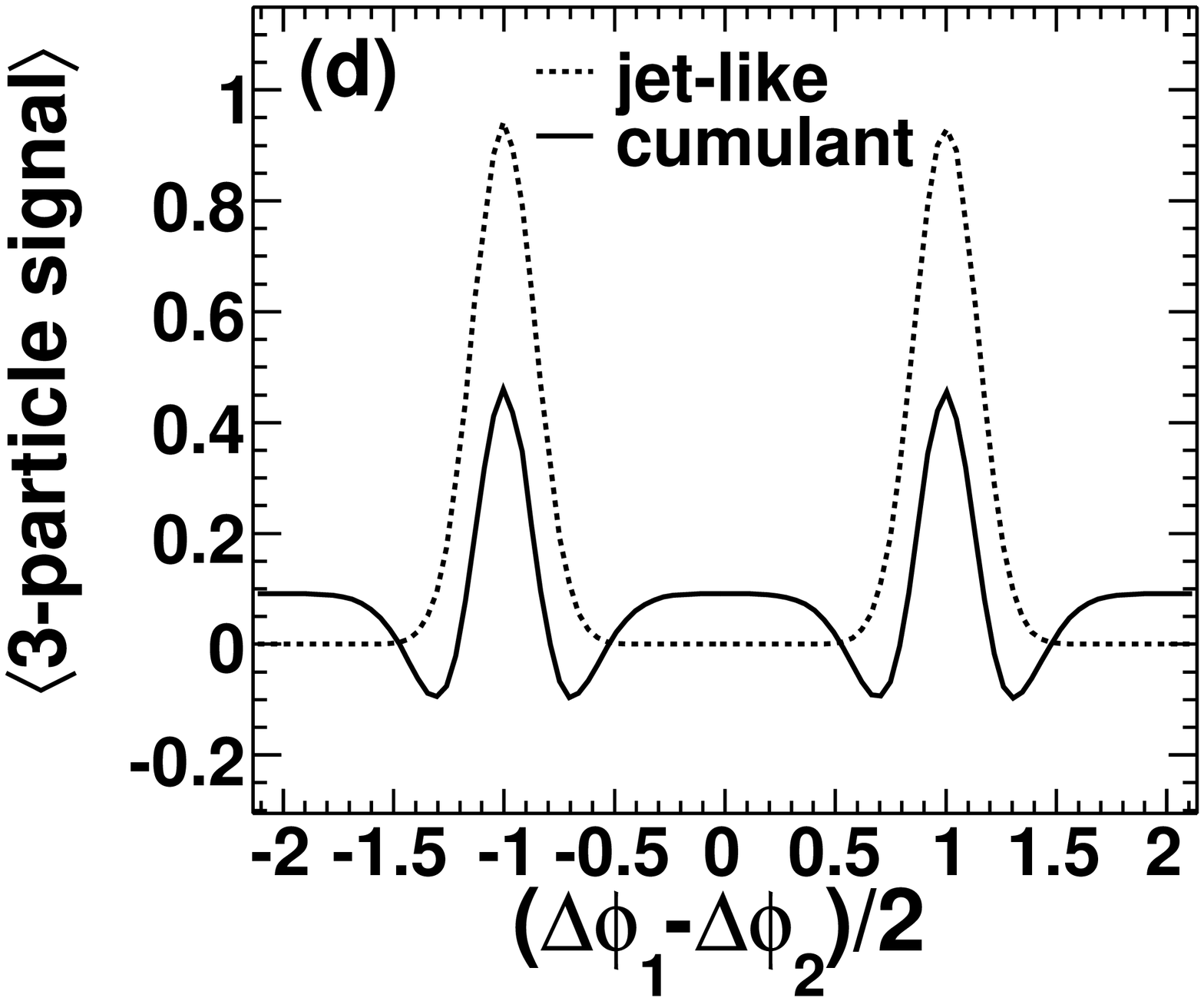,width=0.25\textwidth}
}
\caption{(Color online) Uniform background case (A) with narrow jets and Mach-cones, Eq.~(\ref{eq:caseA}). (a) Two-particle correlation signal atop a uniform background. The dotted line is the estimated background level ($B_1$) to match the signal at the minimum (ZYA1), which is identical to the true background ($\Btrue$). The dash-dotted line is the average of the raw correlation signal ($\rho_1$). (b) Three-particle jet-like correlation after subtracting the background (represented by the dotted horizontal line in the left panel). (c) Three-particle cumulant result, or three-particle jet-like result treating the average raw signal as background [represented by the dash-dotted line in panel (a)]. (d) Comparison between the away-side off-diagonal projections of the three-particle jet-like correlation result in panel (b) and the cumulant result in panel (c). The on-diagonal projection is identical to the off-diagonal projection. The color bars on the right in panels (b) and (c) are three-particle correlation magnitudes.}
\label{fig1}
\end{figure*}

\begin{figure*}[hbt]
\centerline{
\psfig{file=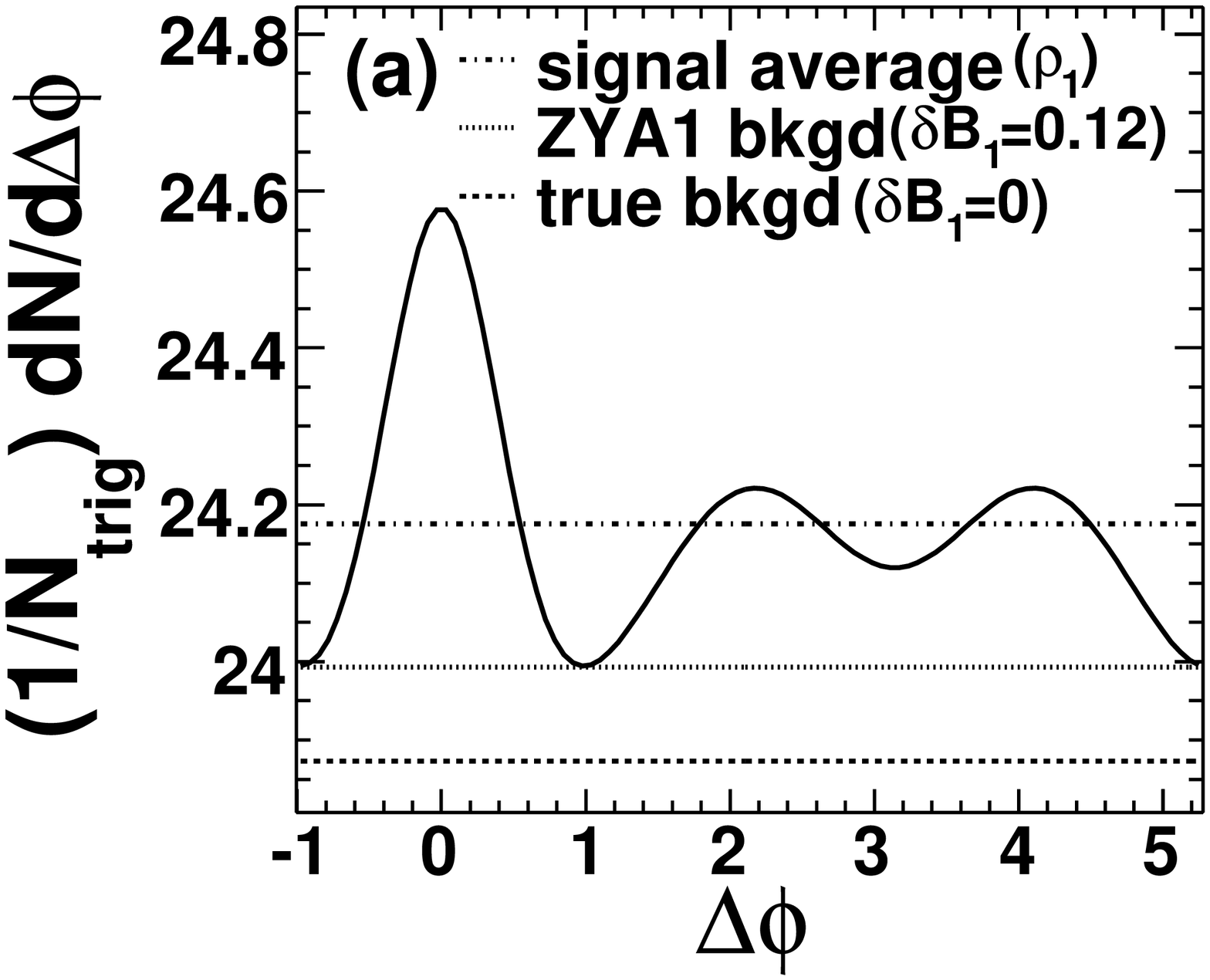,width=0.25\textwidth}
\psfig{file=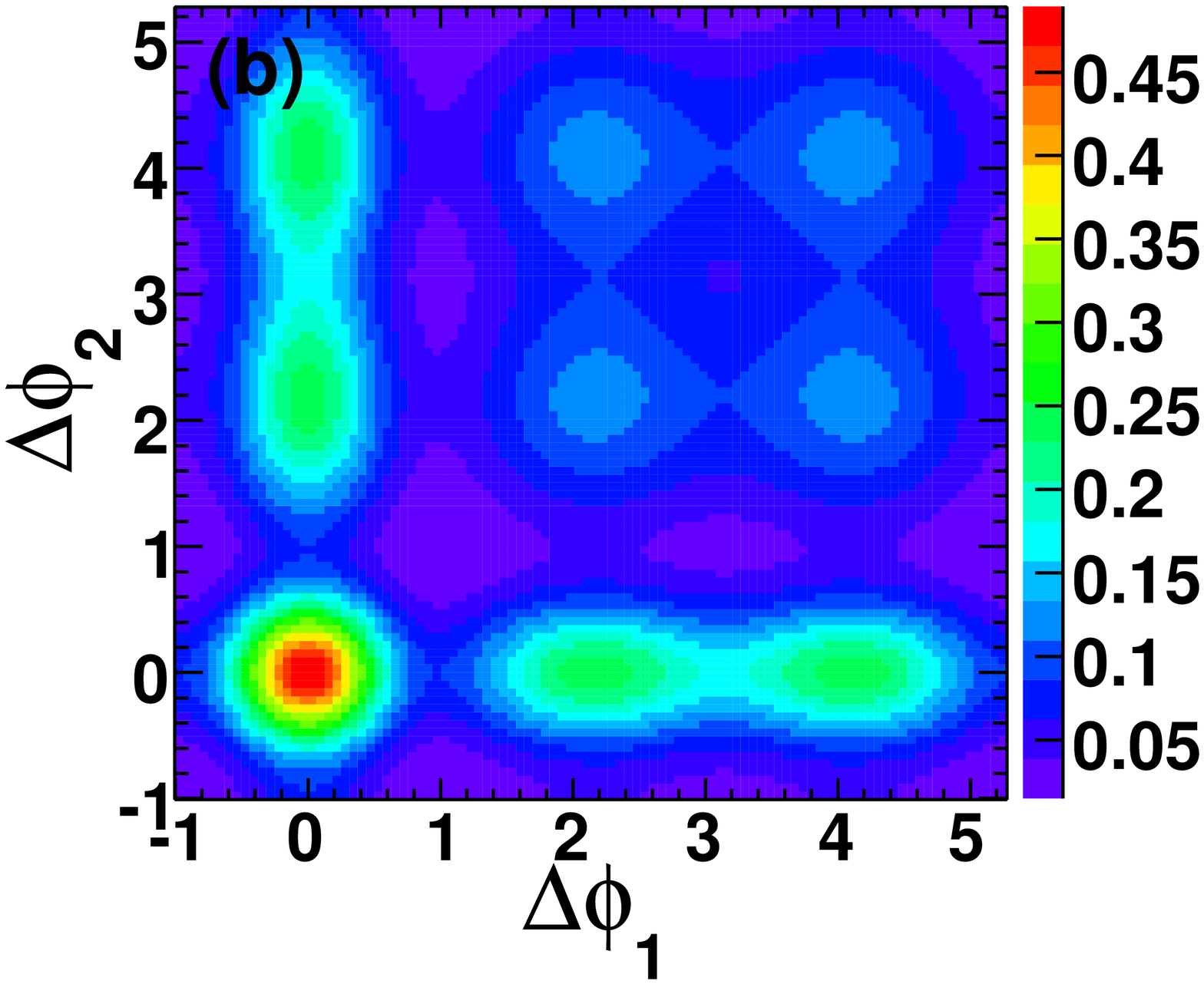,width=0.25\textwidth}
\psfig{file=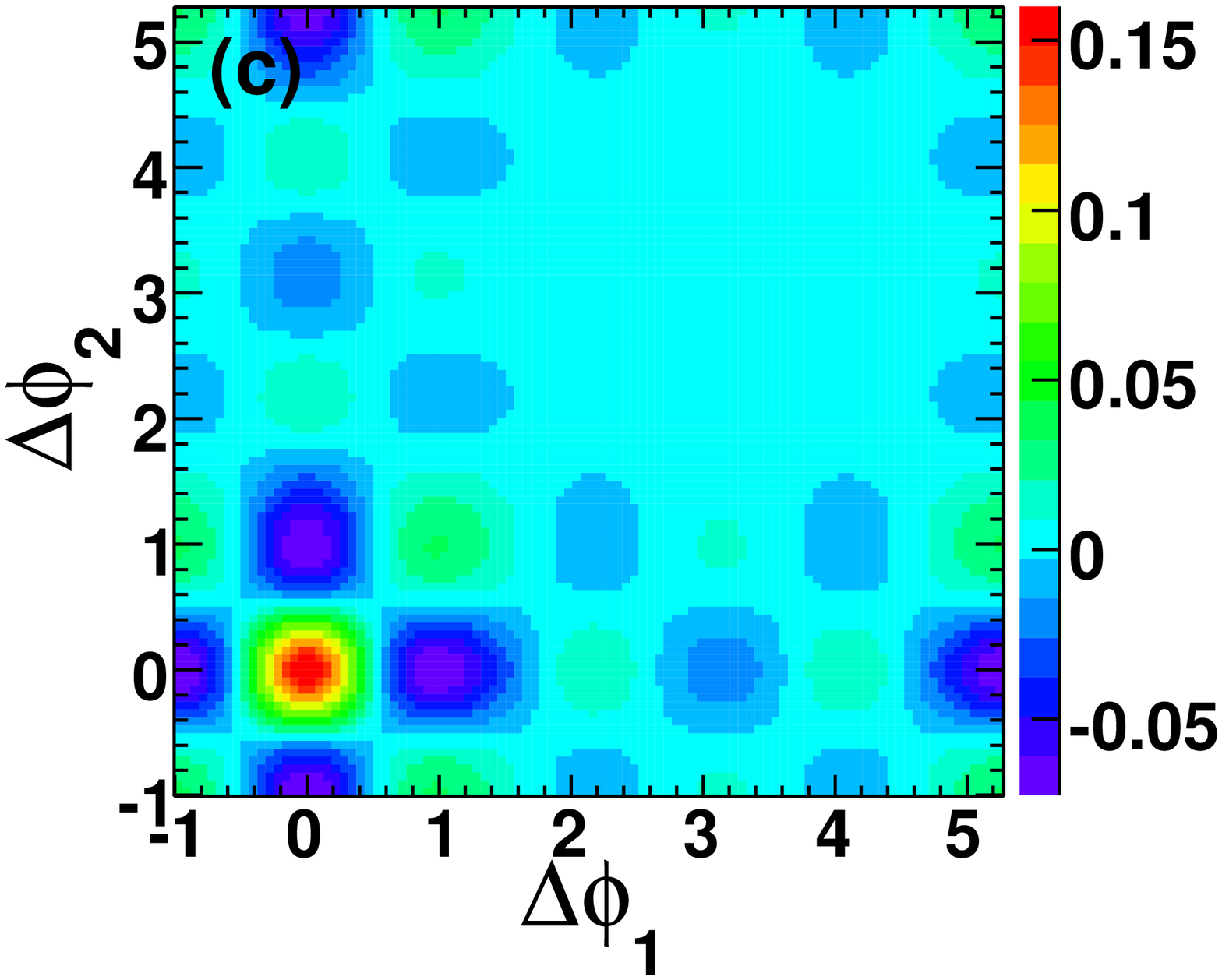,width=0.25\textwidth}
\psfig{file=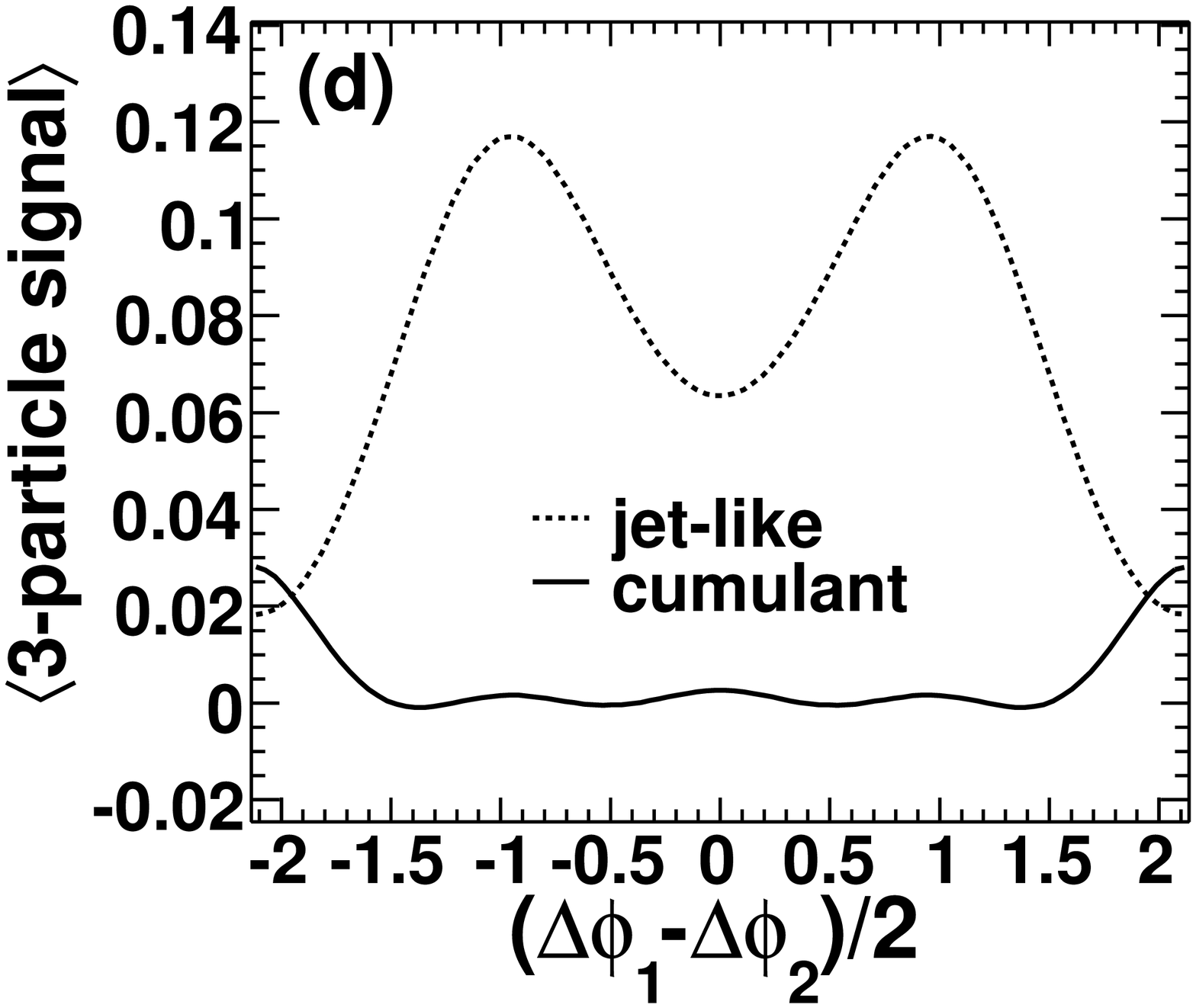,width=0.25\textwidth}
}
\caption{(Color online) Uniform background case (B) with realistic jets and Mach-cones, Eq.~(\ref{eq:caseB}). (a) Two-particle correlation signal atop a uniform background. The dotted line is the estimated background level ($B_1$) to match the signal at the minimum (ZYA1), which is overestimated by $\delta B_1=0.12$ compared to the true background ($\Btrue$) plotted in the dashed line. The dash-dotted line is the average of the raw correlation signal ($\rho_1$). (b) Three-particle jet-like correlation after subtracting the true background [represented by the dashed horizontal line in panel (a)]. (c) Three-particle cumulant result, or three-particle jet-like result treating the average raw signal as background [represented by the dash-dotted line in panel (a)]. (d) Comparison between the away-side off-diagonal projections of the three-particle jet-like correlation result in panel (b) and the cumulant result in panel (c). The on-diagonal projection is identical to the off-diagonal projection. The color bars on the right in panels (b) and (c) are three-particle correlation magnitudes.}
\label{fig2}
\end{figure*}

To give a visual comparison between the two methods, we use a specific example for the jet-correlation signal. We define a jet-like two-particle correlation with a near-side peak and a broad double-hump away-side distribution:
\bw\be\Jhat_2(\dphi)=\frac{N_1}{\sqrt{2\pi}\sigma_1}\exp\left[-\frac{(\dphi)^2}{2\sigma_1^2}\right]+\frac{N_2/2}{\sqrt{2\pi}\sigma_2}\left(\exp\left[-\frac{(\dphi-\pi+\theta)^2}{2\sigma_2^2}\right]+\exp\left[-\frac{(\dphi-\pi-\theta)^2}{2\sigma_2^2}\right]\right).\label{eq:jetGaus}\ee\ew
We suppose this jet-like two-particle correlation is present in every event (i.e., Mach-cone event), and it is atop a large uniform background, $\Btrue=150/2\pi$. We take the cone angle to be $\theta=1$. We study two cases of jet signals: (A) the jets are narrow with $\sigma_1=0.2$ and $\sigma_2=0.2$; (B) we use a realistic jet signal as measured in experiment~\cite{Wang,StarQM05,Star_3part}, with $\sigma_1=0.4$ and $\sigma_2=0.7$. For both cases, we take the numbers of jet-correlated particles to be $N_1=0.7$ and $N_2=1.2$, respectively, for near-side and away-side. These numbers are chosen so that the two-particle jet-correlation signal in the second case corresponds, qualitatively, to the measured one~\cite{Wang,StarQM05,Star_3part}. For both cases the minimum signal strength is at $\dphi\approx 1$; for case (A) the normalized background level is $B_1\approx \Btrue$ (i.e., with a scaling factor of $a=B_1/\rho_1=150/151.9$), and for case (B) the normalized background is $B_1\approx \Btrue+0.12$ (i.e., with a scaling factor of $a=B_1/\rho_1=150.12/151.9$). For easy reference, we list the parameters below:
\be
{\rm (A):}\left\{
\ba{l}
N_1=0.7,N_2=1.2,\sigma_1=0.2,\sigma_2=0.2,\theta=1;\\
\Btrue=150/2\pi,B_1=\Btrue,
\ea
\right.\label{eq:caseA}
\ee
\be
{\rm (B):}\left\{
\ba{l}
N_1=0.7,N_2=1.2,\sigma_1=0.4,\sigma_2=0.7,\theta=1;\\
\Btrue=150/2\pi,B_1=\Btrue+0.12.
\ea
\right.\label{eq:caseB}
\ee
The raw two-particle correlations are shown in Figs.~\ref{fig1}(a) and~\ref{fig2}(a) for case (A) and (B), respectively, and will be described below in detail.

We assume that the three-particle jet-like correlation can be factorized as the product of the two two-particle jet-like correlation functions given by Eq.~(\ref{eq:Jhat3factorize}). The raw three-particle correlation is then given by
\be\ba{rl}
J_3(\dphi_1,\dphi_2)=&\Jhat_2(\dphi_1)\Jhat_2(\dphi_2)+\Btrue^2\\
&+\Jhat_2(\dphi_1)\Btrue+\Jhat_2(\dphi_2)\Btrue\\
=&\left[\Jhat_2(\dphi_1)+\Btrue\right]\left[\Jhat_2(\dphi_2)+\Btrue\right].
\ea\label{eq18}\ee
This raw three-particle correlation is needed as the starting point of any data analysis. One can extract the three-particle jet-correlation from the raw signal by making different assumptions of the underlying background level, which is the essential difference between the jet-correlation method and the cumulant method.

The raw two-particle correlation for case (A) is shown in Fig.~\ref{fig1}(a). The dotted line shows the real background level, $\Btrue$, (or the normalized background level $B_1$, which equals to $\Btrue$). The dash-dotted line shows the single particle density, $\rho_1$, used in the cumulant method. The three-particle correlation signal after subtraction of the real background (i.e., $\Btrue$) is shown in Fig.~\ref{fig1}(b). This is the genuine three-particle correlation function that was initially put in, which shows the distinctive Mach-cone structure. The three-particle cumulant result is shown in Fig.~\ref{fig1}(c). This is equivalent to what the three-particle jet-correlation analysis would yield by subtracting an overestimated background level of $B_1=\rho_1$, given by Eq.~(\ref{eq15}) where $\mean{\Jhat_2}=\rho_1-\Btrue$. The Mach-cone structure is partially preserved in the three-particle cumulant result. The negative strips are produced by the over-subtraction of the average signal particle density, which is larger than the true background level. More direct comparison between the two methods can be achieved by projections of the corresponding three-particle correlation signals. Figure~\ref{fig1}(d) compares the projections along the off-diagonal axis on the away-side, $\dphi_{1,2}>1$. The away-side on-diagonal projections are identical because the input to the simulation is symmetric. Both methods show a clear Mach-cone signal at the input cone angle ($\theta=1$). The structure from the cumulant result is more complex than that from the jet-correlation method. Our cumulant results are consistent with studies in~\cite{cumulant} where strong and narrow Mach-cone signal is also used.

The raw two-particle correlation for case (B) is shown in Fig.~\ref{fig2}(a). The dashed line shows the real background level $\Btrue$. The dotted line shows the normalized background level $B_1$ by ZYA1, such that the jet signal is zero at minimum. The dash-dotted line shows the single particle density, $\rho_1$, used in the cumulant method. The three-particle jet-correlation signal after subtraction of the real background level (i.e., $\Btrue$) is shown in Fig.~\ref{fig2}(b). Again, this is the genuine three-particle jet-correlation function that was initially put in, $\Jhat_3(\dphi_1,\dphi_2)=\Jhat_2(\dphi_1)\Jhat_2(\dphi_2)$. Figure~\ref{fig2}(c) shows the three-particle cumulant result. The structure of the cumulant result is quite complex and different from the three-particle jet-correlation result. This is more clearly shown in the away-side off-diagonal projections in Fig.~\ref{fig2}(d). The complex structure in cumulant is due to subtraction of the average single particle density, which is larger than the true background level. Unlike the simple case (A) where the jet peaks are narrow and well-confined, the broad jet peaks and over-subtraction of the background level in case (B) create the complex structure in the cumulant result. The distinctive Mach-cone structure on the away-side that is initially put in is hardly observable in the cumulant result. This is because the away-side two-particle raw correlation as shown in Fig.~\ref{fig2}(a), qualitatively similar to that measured in real data~\cite{Wang,StarQM05,Star_3part}, is roughly flat and happens to have similar magnitude as the average single particle density. After subtraction of the single particle density, the away-side two-particle correlation strength is more or less subtracted away and can hardly show up in the final three-particle cumulant. 

\begin{figure*}[hbt]
\centerline{
\psfig{file=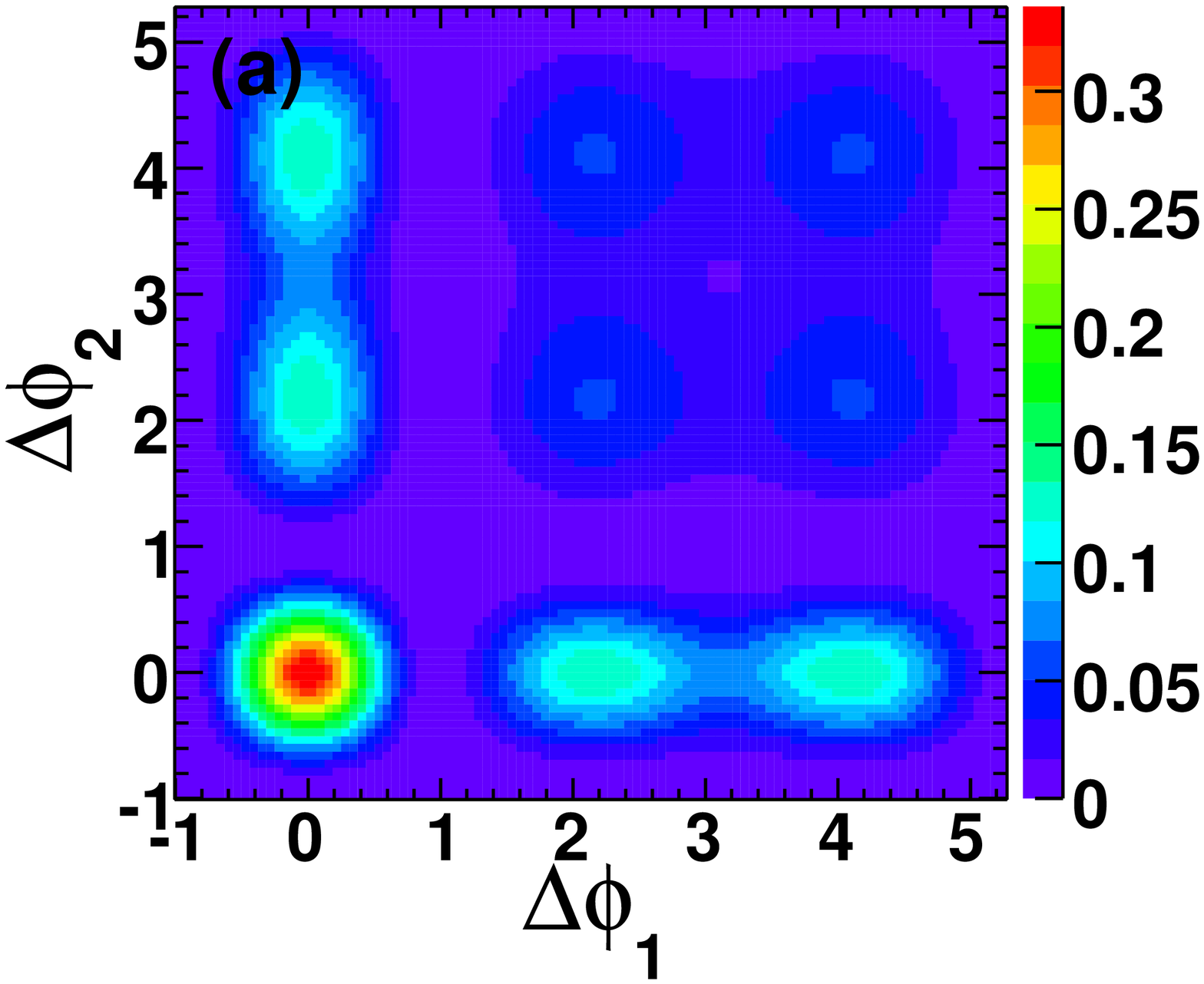,width=0.25\textwidth}
\psfig{file=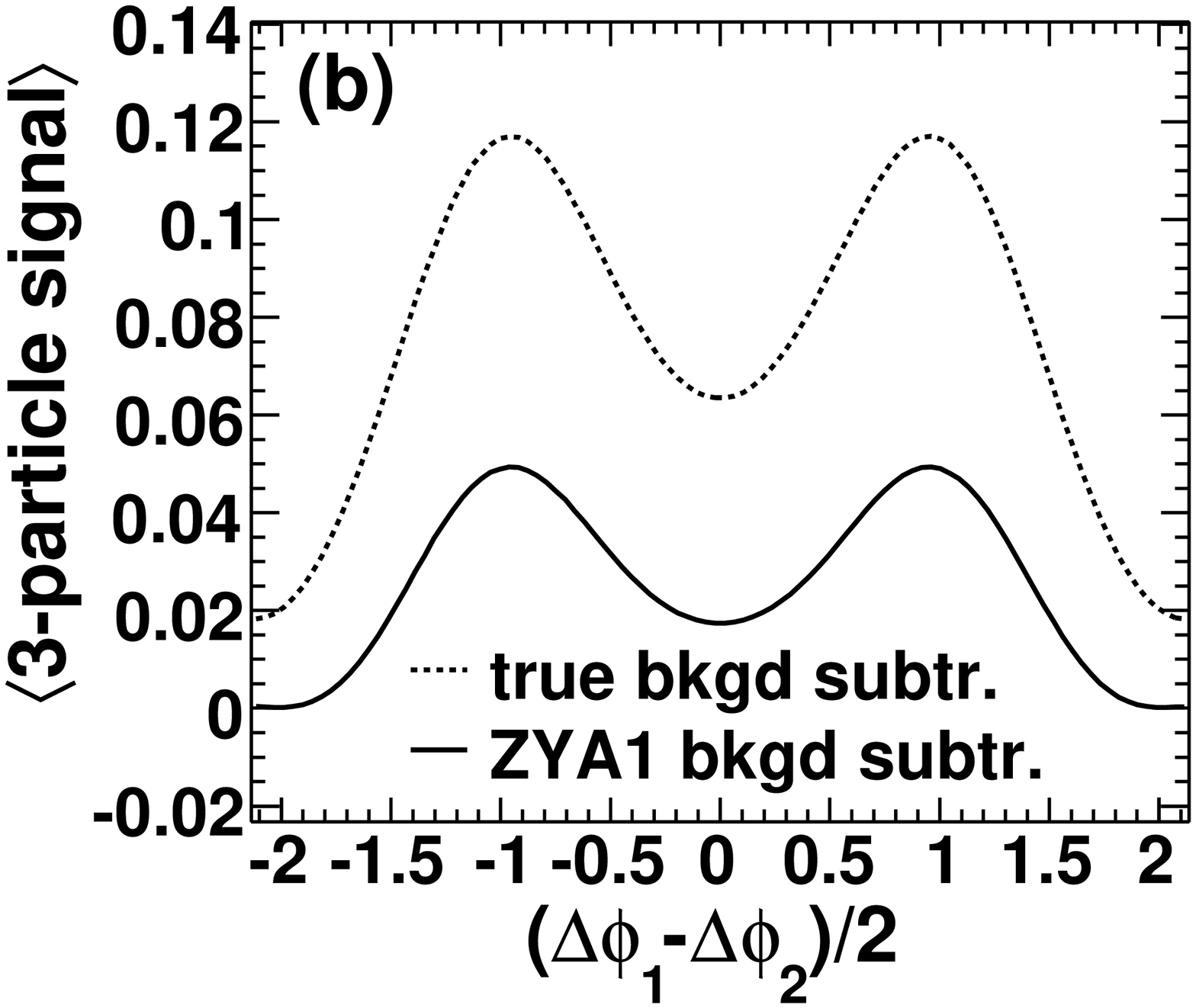,width=0.25\textwidth}
\psfig{file=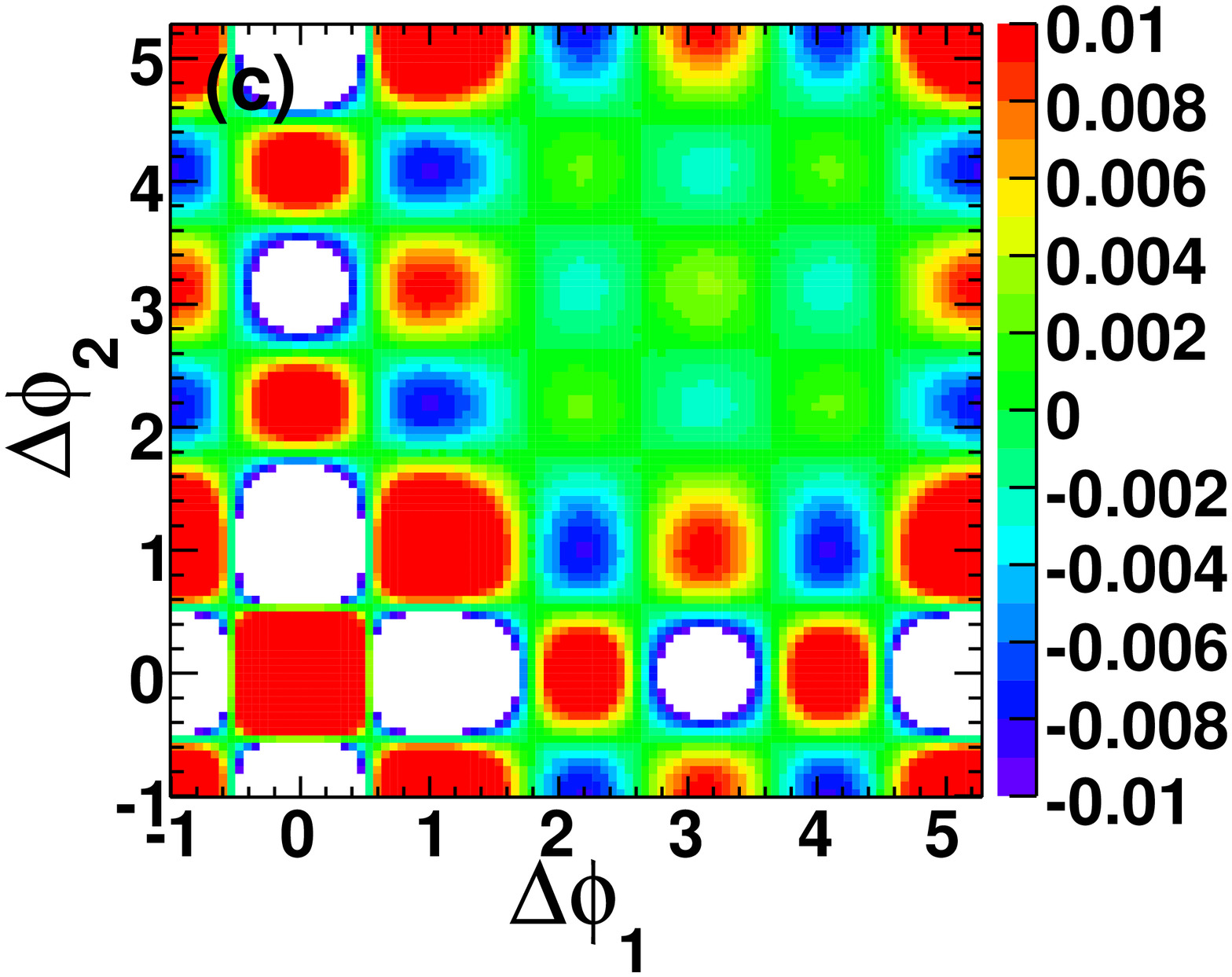,width=0.25\textwidth}
}
\caption{(Color online) Uniform background case (B) with realistic jets and Mach-cones as in Fig.~\ref{fig2}. (a) Three-particle jet-like correlation with background normalized ($B_1=\Btrue+\delta B_1$, $\delta B_1=0.12$) to the signal at $\dphi=1$ [represented by the dotted horizontal line in Fig.~\ref{fig2}(a)]. Note the Mach-cone structure is still observable. (b) Away-side off-diagonal projections of the three-particle jet-like correlation result subtracting the true (input) background [shown in Fig.~\ref{fig2}(b)] and that subtracting ZYA1 background [shown in panel (a)]. The dashed curve is as same as that in Fig.~\ref{fig2}(d). (c) Zoomed-in version of Fig.~\ref{fig2}(c). The color bars on the right in panels (a) and (c) are three-particle correlation magnitudes.}
\label{fig3}
\end{figure*}

The three-particle jet-correlation signal after subtraction of the normalized background level (i.e., $B_1$) is shown in the left panel of Fig.~\ref{fig3}(a). This would be the three-particle jet-correlation function from data analysis using ZYA1 or ZYAM background normalization scheme. Since the background is overestimated, the obtained three-particle jet-correlation signal is lower than the true signal, but the structure of the genuine three-particle jet-correlation (i.e., the Mach-cone structure) is preserved. This is more clearly seen in the away-side off-diagonal projections shown in Fig.~\ref{fig3}(b) where the dashed histogram is projected from the real three-particle correlation in Fig.~\ref{fig2}(b) and the solid histogram is projected from the three-particle correlation result obtained with ZYA1 background normalization [i.e., from Fig.~\ref{fig3}(a)]. In fact, the three-particle jet-correlation signal with the overestimated background, $B_1>\Btrue$, is given by 
\bea
\Jhat_3(\dphi_1,\dphi_2)&=&\left[\Jhat_2(\dphi_1)-\delta B_1\right]\left[\Jhat_2(\dphi_2)-\delta B_1\right]\label{eq19}\\
&&({\rm where}\;\; \delta B_1=B_1-\Btrue).\nonumber
\eea
Note the similarity between Eqs.~(\ref{eq19}) and~(\ref{eq15}): if the uniform background levels are chosen to be equal, then the resultant three-particle jet-correlation and three-particle cumulant are identical.

On the other hand, if one zooms into the cumulant result in Fig.~\ref{fig2}(c), now shown in Fig.~\ref{fig3}(c), one may also see the Mach-cone structure due to the double-hump in the away-side two-particle correlation.
%The structure due to the double-hump in the away-side two-particle correlation, which resembles the Mach-cone structure, is visible on this fine level of the current analytical demonstration of the cumulant result. However, it would be hard to observe in a real data analysis due to the other present complicated structures and limited statistics.
However, it would be extremely difficult to distinguish it from the many other and much larger peaks in the cumulant result.

%%%%%%%%%%%%%%%%%%%%%%%%%%%%%%%%%%%%%%%%%%%%%%%%%%%%%%%%%%%%%%%%%%%%%%
\subsection{Anisotropic flow without jet-correlation}

In heavy-ion collisions, particles are correlated to reaction plane due to the hydrodynamic type of collective flow of the bulk medium and the anisotropic overlap region between the colliding nuclei. The trigger particle emission is also correlated to reaction plane due to, not so much of hydrodynamic flow, but the path-length dependent energy loss of high $\pt$ particles in the medium that is initially anisotropic. This reaction plane correlation, expressed in harmonics up to the fourth order, is given by
\be B_2(\dphi)=B_1\left[1+2\vtrig v_2\cos(2\dphi)+2\vvtrig v_4\cos(4\dphi)\right],\label{eq:B2_flow}\ee
where $\vtrig$ and $v_2$ are the elliptic flow parameters of the trigger and associated particles, respectively. Likewise $\vvtrig$ and $v_4$ are the respective fourth harmonic coefficients. The two-particle jet-correlation signal of Eq.~(\ref{eq5}) is then,
\be
\ba{ll}
\Jhat_2(\dphi)=&J_2(\dphi)\\
&-B_1\left[1+2\vtrig v_2\cos(2\dphi)+2\vvtrig v_4\cos(4\dphi)\right].
\ea
\label{eq:Jhat2_flow}
\ee

\begin{figure*}[hbt]
\centerline{
\psfig{file=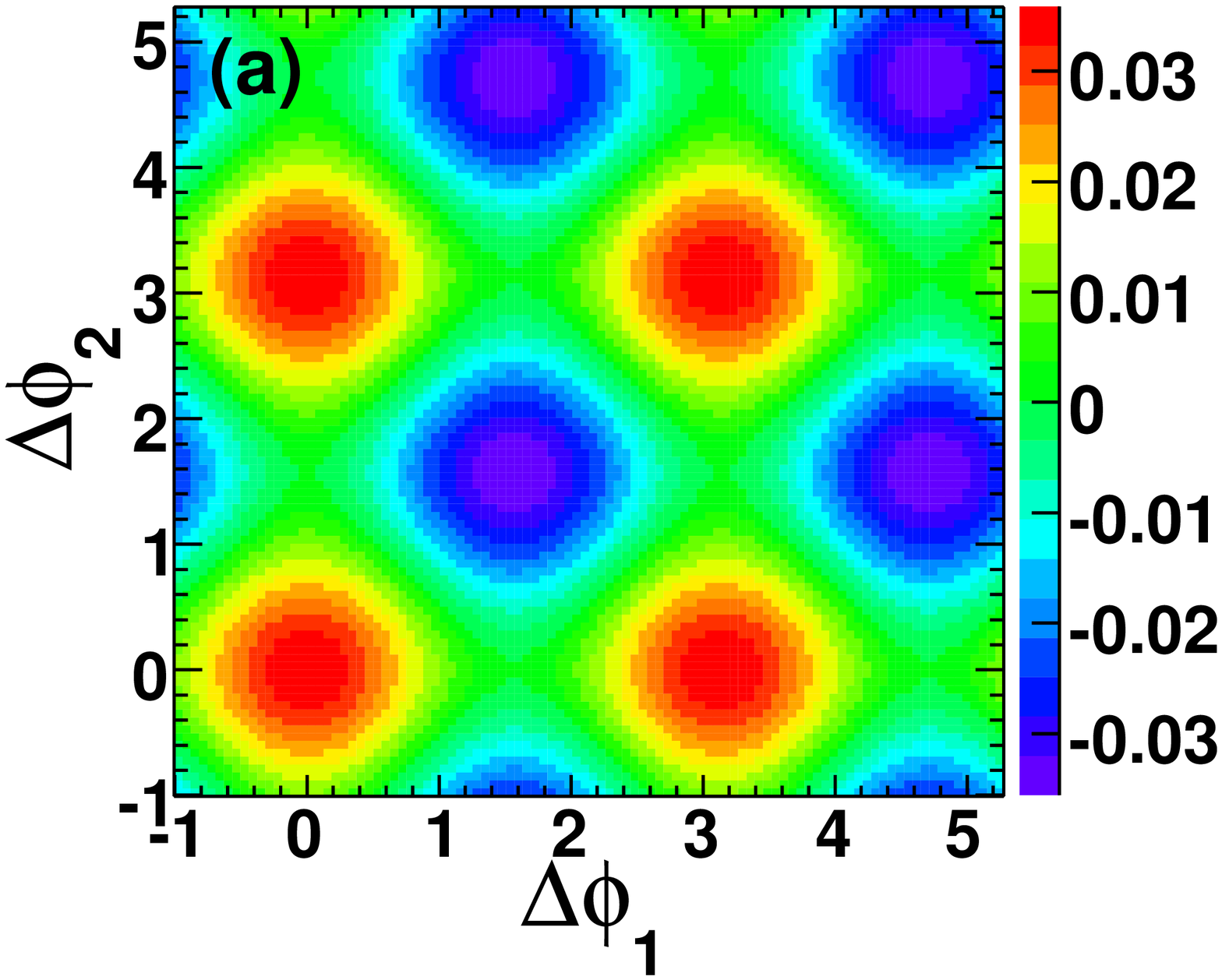,width=0.25\textwidth}
\psfig{file=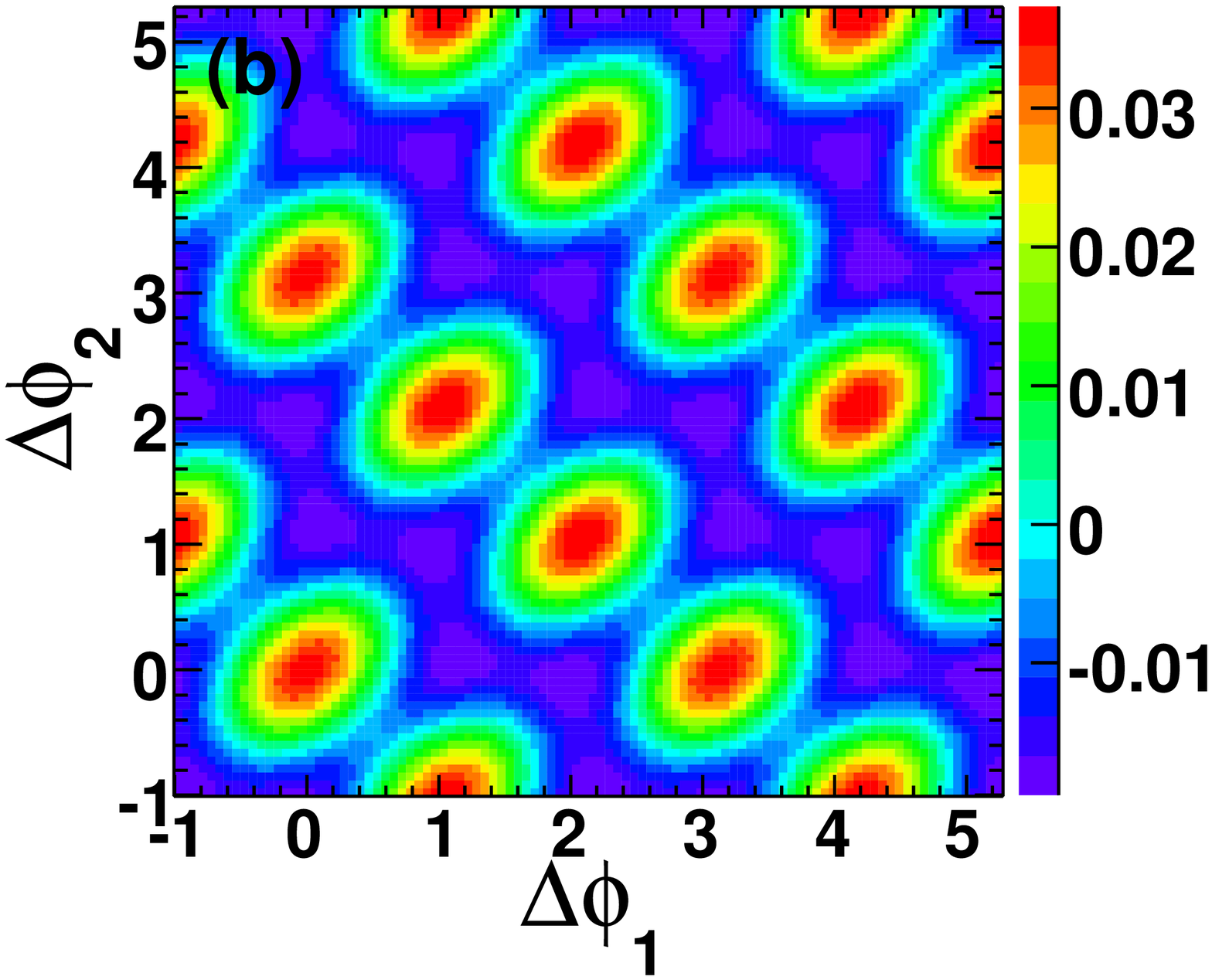,width=0.25\textwidth}
\psfig{file=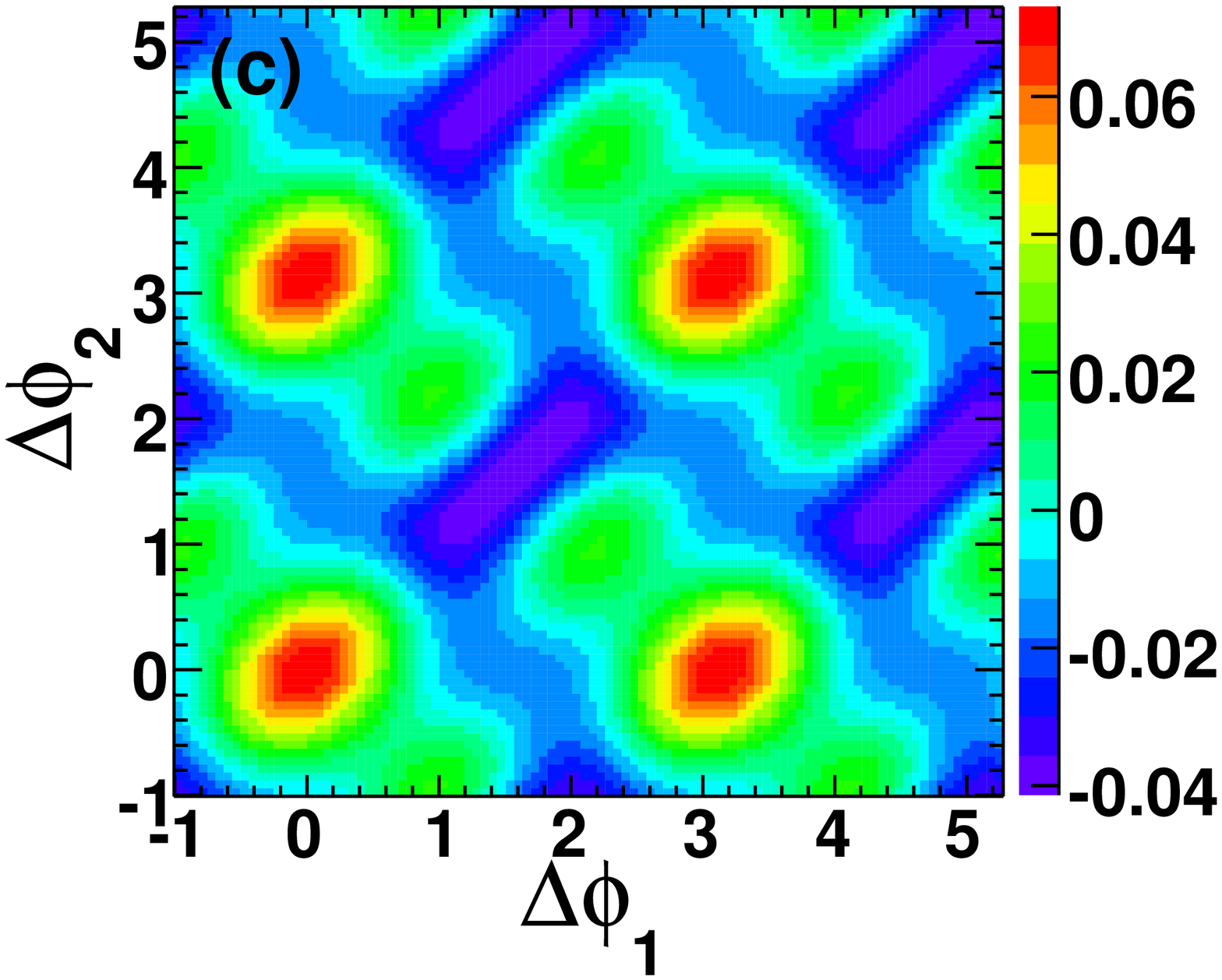,width=0.25\textwidth}
\psfig{file=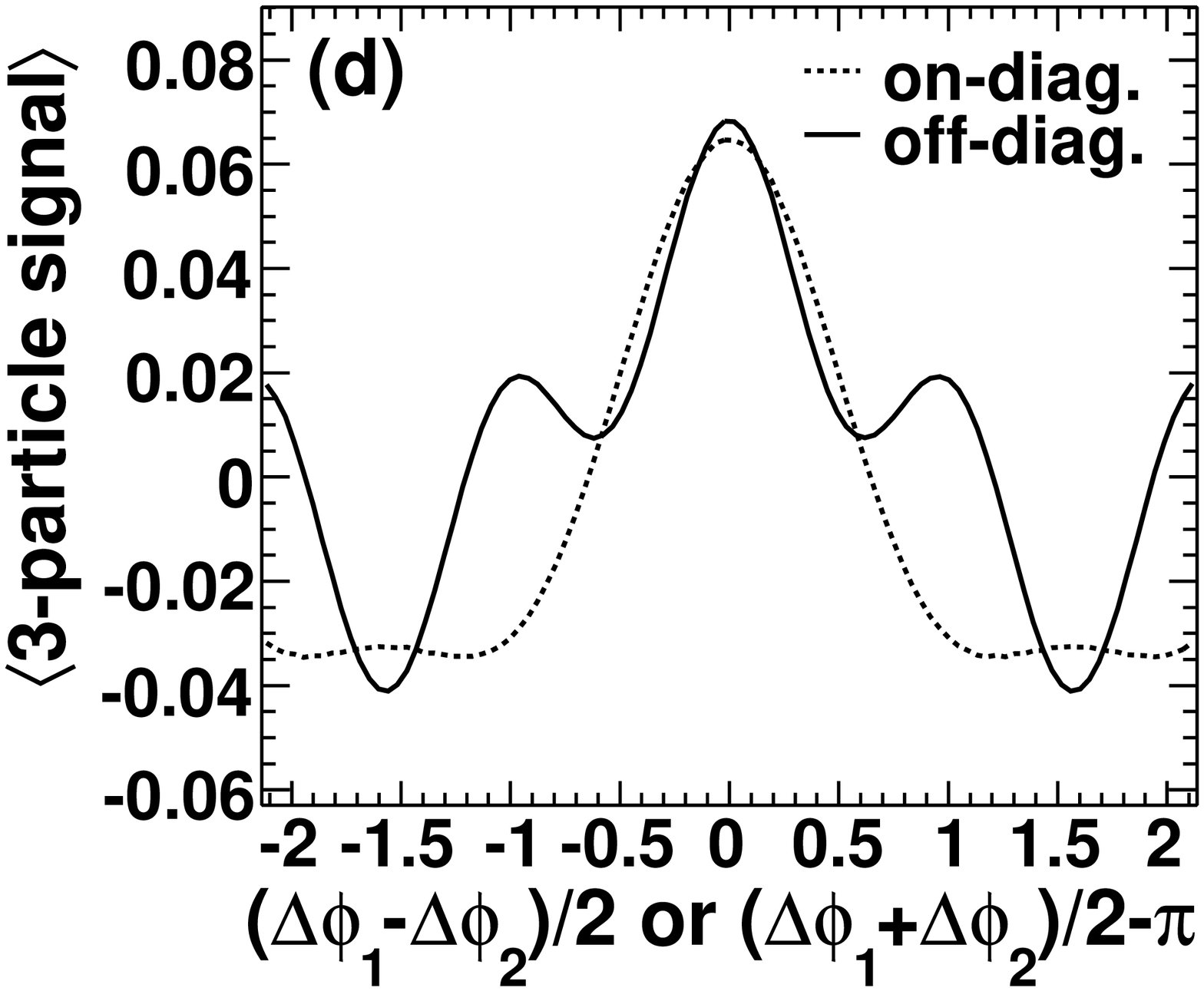,width=0.25\textwidth}
}
\caption{(Color online) Lab-frame three-particle cumulant from pure anisotropic flow correlation, Eq.~(\ref{eq:flow_value}). (a) Non-Poisson statistics induced cumulant with typical $\mean{B_1^2}-\mean{B_1}^2=0.1\mean{B_1}$. (b) Irreducible flow correlation terms (where $\mean{B_1^2}=\mean{B_1}^2$ is taken). (c) Total cumulant given by Eq.~\ref{eq26} [sum of panel (a) and (b)]. (d) Away-side on-diagonal and off-diagonal projections of the total cumulant in panel (c). The color bars on the right in panels (a), (b), and (c) are three-particle correlation magnitudes.}
\label{fig4}
\end{figure*}

There are two combinatorial backgrounds to three-particle jet-correlation. One is that of a correlated trigger-associated pair combined with a random background particle. This background can be obtained by the product of the two-particle jet-correlation signal with the underlying background particle, namely $\Jhat_2(\dphi_1)B_2(\dphi_2)+\Jhat_2(\dphi_2)B_2(\dphi_1)$. The other combinatorial background is due to the trigger particle combined with two random particles from the underlying background. This background term normalized per trigger particle, considering only the anisotropic flow correlation, is given by~\cite{method}
\bw\be
B_3(\dphi_1,\dphi_2)=\mean{B_1^2}\left(
\ba{ll}
1&+2\vtrig\vone\cos(2\dphi_1)+2\vtrig\vtwo\cos(2\dphi_2)+2\vone\vtwo\cos2(\dphi_1-\dphi_2)\\
 &+2\vvtrig\vvone\cos(4\dphi_1)+2\vvtrig\vvtwo\cos(4\dphi_2)+2\vvone\vvtwo\cos4(\dphi_1-\dphi_2)\\
 &+2\vtrig\vone\vvtwo\cos2(\dphi_1-2\dphi_2)+2\vtrig\vtwo\vvone\cos2(2\dphi_1-\dphi_2)\\
 &+2\vone\vtwo\vvtrig\cos2(\dphi_1+\dphi_2)
\ea
\right),\label{eq22}
\ee\ew
where $\mean{B_1^2}$ is the per trigger associated pair density which is not necessarily equal to $\mean{B_1}^2\equiv B_1^2$ (i.e., non-Poisson statistics). In Eq.~(\ref{eq22}) and hereafter, $\vone$ and $\vtwo$ are elliptic flow parameters of the two associated particles, and $\vvone$ and $\vvtwo$ are the respective fourth harmonic coefficients. 
%We note that in real collision data the background particles possess not only anisotropic flow correlation, but also other correlations that are unrelated to the trigger particle, such as jet-correlations due to jets other than the one selected by the trigger particle~\cite{estruct}. As mentioned, we do not consider these correlations in this paper because they are not important for the purpose of our study. 
The final three-particle jet-correlation function is given by Eq.~(\ref{eq6a}) or Eq.~(\ref{eq6b}).
%\be\ba{ll}
%\Jhat_3(\dphi_1,\dphi_2)&=J_3(\dphi_1,\dphi_2)-B_3(\dphi_1,\dphi_2)-\\
%&\Jhat_2(\dphi_1)B_2(\dphi_2)-\Jhat_2(\dphi_2)B_2(\dphi_1)
%\ea\label{eq23a}\ee
%Or alternatively
%\be\ba{ll}
%\Jhat_3(\dphi_1,\dphi_2)=&J_3(\dphi_1,\dphi_2)-B_3(\dphi_1,\dphi_2)-\\
%&J_2(\dphi_1)B_2(\dphi_2)-J_2(\dphi_2)B_2(\dphi_1)+\\
%&2B_2(\dphi_1)B_2(\dphi_2).
%\ea\label{eq23b}\ee

It is interesting to consider only anisotropic flow correlation, no jet-like correlation. Obviously, if no jet-like correlation is present, the three-particle jet-correlation will give zero signal as there will be no two-particle jet-correlation signal to start with, $\Jhat_2(\dphi)=0$. However, the three-particle cumulant will still yield nonzero result, as we demonstrate below. 

With only anisotropic flow correlation present, the raw one-, two-, and three-particle lab-frame cumulants are simply given by the background, $\rho_1=B_1$, $\rho_2(\dphi)=B_2(\dphi)$, and $\rho_3(\dphi_1,\dphi_2)=B_3(\dphi_1,\dphi_2)$. Thus the three-particle cumulant from Eq.~(\ref{eq4a}) is
\be\ba{ll}
\rhohat_3(\dphi_1,\dphi_2)=&B_3(\dphi_1,\dphi_2)-B_2(\dphi_1-\dphi_2)\\
&-B_2(\dphi_1)B_1-B_2(\dphi_2)B_1+2B_1^2.
\ea\label{eq24}\ee
Here $B_2(\dphi)$ is given by Eq.~(\ref{eq:B2_flow}), $B_3(\dphi_1,\dphi_2)$ by Eq.~(\ref{eq22}), and $B_2(\dphi_1-\dphi_2)$ is the two-particle flow correlation normalized per trigger particle, and is given by
\bw
\be B_2(\dphi_1-\dphi_2)=\mean{B_1^2}\left[1+2\vone\vtwo\cos2(\dphi_1-\dphi_2)+2\vvone\vvtwo\cos4(\dphi_1-\dphi_2)\right].\label{eq25}\ee
With simple algebra, we obtain
\be\ba{ll}
\rhohat_3(\dphi_1,\dphi_2)=&2\left(\mean{B_1^2}-\mean{B_1}^2\right)\left[\vtrig\vone\cos(2\dphi_1)+\vvtrig\vvone\cos(4\dphi_1)+\vtrig\vtwo\cos(2\dphi_2)+\vvtrig\vvtwo\cos(4\dphi_2)\right]\\
&+2\mean{B_1^2}\left[\vtrig\vone\vvtwo\cos2(\dphi_1-2\dphi_2)+\vtrig\vtwo\vvone\cos2(2\dphi_1-\dphi_2)+\vone\vtwo\vvtrig\cos2(\dphi_1+\dphi_2)\right].\label{eq26}
\ea\ee
\ew
With no jet correlation, only anisotropic flow correlation present, the three-particle cumulant gives non-zero correlation result. The three-particle lab-frame cumulant measures something different from simple jet-correlation; it measures three-particle correlation regardless of the nature of the underlying physics. With Poisson statistics, $\mean{B_1^2}=\mean{B_1}^2$, the first square-bracket term in the r.h.s.~of Eq.~(\ref{eq26}) drops and what remains are the irreducible flow correlation terms (in the second square bracket). Although these terms are on the order of $v_2^4$, their magnitudes can be still sizable due to the large background level of $B_1^2$ in relativistic heavy-ion collisions. With non-Poisson statistics, the cumulant is even more complicated. To give a visual impression of the cumulant from pure anisotropic flow, we plot Eq.~(\ref{eq26}) with typical flow magnitudes:
\be
\vtrig=7.5\%,\vone=\vtwo=5\%,v_4=v_2^2.
\label{eq:flow_value}
\ee
The result is shown in Fig.~\ref{fig4}. Panel (a) shows the non-Poisson statistics induced cumulant with typical 
\be\mean{B_1^2}-\mean{B_1}^2=0.1\mean{B_1}.\ee
Panel (b) shows the irreducible flow correlation terms where we have taken $\mean{B_1^2}=\mean{B_1}^2$. Panel (c) shows the total cumulant. Panel (d) shows the away-side on-diagonal and off-diagonal projections of the total cumulant. As seen, pure flow gives non-zero cumulant with complex structures.

%%%%%%%%%%%%%%%%%%%%%%%%%%%%%%%%%%%%%%%%%%%%%%%%%%%%%%%%%%%%%%%%%%%%%%
\subsection{Jet-correlation with anisotropic flow background}

We now turn our attention to the situation where both jet-correlation and anisotropic flow are present. For the sake of simplicity we shall assume Poisson statistics 
\be\mean{B_1^2}=\mean{B_1}^2\equiv B_1^2.\ee 

If the true background level and the anisotropic flow magnitude are precisely known, then all the background terms are determined, and the final jet-like three-particle correlation signal from the jet-correlation method recovers the true input signal of Eq.~(\ref{eq:jetGaus}), the same as that shown in Fig.~\ref{fig2}(b). However, experimentally the true background level is not known a priori, and thus background normalization schemes, such as ZYA1 or ZYAM, have to be employed. We shall examine the effect of an imperfect background estimate in the jet-correlation method. We will not discuss the effect of uncertainties in the anisotropic flow magnitudes as they have been extensively discussed in Ref.~\cite{method}. Moreover, systematic uncertainties on the flow measurement can be reliably accessed in experiment.

From Eqs.~(\ref{eq6a}), (\ref{eq:B2_flow}), (\ref{eq:Jhat2_flow}), and (\ref{eq22}), an uncertainty $\delta B_1$ in the background level $B_1$ introduces a change in the three-particle correlation signal from the jet-correlation method:
\bw
\bea
\delta\Jhat_3	%(\dphi_1,\dphi_2)
%&=&-[2B_1\delta B_1+(\delta B_1)^2]\left(
%\ba{ll}
%1&+2\vtrig\vone\cos(2\dphi_1)+2\vtrig\vtwo\cos(2\dphi_2)+2\vone\vtwo\cos2(\dphi_1-\dphi_2)\\
% &+2\vvtrig\vvone\cos(4\dphi_1)+2\vvtrig\vvtwo\cos(4\dphi_2)+2\vvone\vvtwo\cos4(\dphi_1-\dphi_2)\\
% &+2\vtrig\vone\vvtwo\cos2(\dphi_1-2\dphi_2)+2\vtrig\vtwo\vvone\cos2(2\dphi_1-\dphi_2)\\
% &+2\vone\vtwo\vvtrig\cos2(\dphi_1+\dphi_2)
%\ea
%\right)\nonumber\\
%&&-\delta B_1\left[1+2\vtrig\vtwo\cos(2\dphi_2)+2\vvtrig\vvtwo\cos(4\dphi_2)\right]\left(\Jhat_2(\dphi_1)+B_1\left[1+2\vtrig\vone\cos(2\dphi_1)+2\vvtrig\vvone\cos(4\dphi_1)\right]\right)\nonumber\\
%&&-\delta B_1\left[1+2\vtrig\vone\cos(2\dphi_1)+2\vvtrig\vvone\cos(4\dphi_1)\right]\left(\Jhat_2(\dphi_2)+B_1\left[1+2\vtrig\vtwo\cos(2\dphi_2)+2\vvtrig\vvtwo\cos(4\dphi_2)\right]\right)\nonumber\\
%&&+2[2B_1\delta B_1+(\delta B_1)^2]\left[1+2\vtrig\vtwo\cos(2\dphi_2)+2\vvtrig\vvtwo\cos(4\dphi_2)\right]\left[1+2\vtrig\vone\cos(2\dphi_1)+2\vvtrig\vvone\cos(4\dphi_1)\right]\nonumber\\
&=&-[2B_1\delta B_1+(\delta B_1)^2]\left(
\ba{ll}
1&+2\vtrig\vone\cos(2\dphi_1)+2\vtrig\vtwo\cos(2\dphi_2)+2\vone\vtwo\cos2(\dphi_1-\dphi_2)\\
 &+2\vvtrig\vvone\cos(4\dphi_1)+2\vvtrig\vvtwo\cos(4\dphi_2)+2\vvone\vvtwo\cos4(\dphi_1-\dphi_2)\\
 &+2\vtrig\vone\vvtwo\cos2(\dphi_1-2\dphi_2)+2\vtrig\vtwo\vvone\cos2(2\dphi_1-\dphi_2)\\
 &+2\vone\vtwo\vvtrig\cos2(\dphi_1+\dphi_2)
\ea
\right)\nonumber\\
&&-\delta B_1\Jhat_2(\dphi_1)\left[1+2\vtrig\vtwo\cos(2\dphi_2)+2\vvtrig\vvtwo\cos(4\dphi_2)\right]\nonumber\\
&&-\delta B_1\Jhat_2(\dphi_2)\left[1+2\vtrig\vone\cos(2\dphi_1)+2\vvtrig\vvone\cos(4\dphi_1)\right]\nonumber\\
&&+2[B_1\delta B_1+(\delta B_1)^2]
%\left(
%\ba{ll}
%1&+2\vtrig\vone\cos(2\dphi_1)+2\vtrig\vtwo\cos(2\dphi_2)\\
% &+2\vvtrig\vvone\cos(4\dphi_1)+2\vvtrig\vvtwo\cos(4\dphi_2)\\
% &+4(\vtrig)^2\vone\vtwo\cos(2\dphi_1)\cos(2\dphi_2)+4\vtrig\vvtrig\vvone\vtwo\cos(4\dphi_1)\cos(2\dphi_2)\\
% &+4\vtrig\vvtrig\vone\vvtwo\cos(2\dphi_1)\cos(4\dphi_2)+4(\vvtrig)^2\vvone\vvtwo\cos(4\dphi_1)\cos(4\dphi_2)
%\ea
%\right)
\left[1+2\vtrig\vtwo\cos(2\dphi_2)+2\vvtrig\vvtwo\cos(4\dphi_2)\right]\left[1+2\vtrig\vone\cos(2\dphi_1)+2\vvtrig\vvone\cos(4\dphi_1)\right]
\nonumber\\
&=&-\delta B_1\Jhat_2(\dphi_1)\left[1+2\vtrig\vtwo\cos(2\dphi_2)+2\vvtrig\vvtwo\cos(4\dphi_2)\right]\nonumber\\
&&-\delta B_1\Jhat_2(\dphi_2)\left[1+2\vtrig\vone\cos(2\dphi_1)+2\vvtrig\vvone\cos(4\dphi_1)\right]\nonumber\\
&&+(\delta B_1)^2\left[1+2\vtrig\vtwo\cos(2\dphi_2)+2\vvtrig\vvtwo\cos(4\dphi_2)\right]\left[1+2\vtrig\vone\cos(2\dphi_1)+2\vvtrig\vvone\cos(4\dphi_1)\right]\nonumber\\
&&-[2B_1\delta B_1+(\delta B_1)^2]\left(
\ba{l}
2\vone\vtwo\cos2(\dphi_1-\dphi_2)+2\vvone\vvtwo\cos4(\dphi_1-\dphi_2)+\\
2\vtrig\vone\vvtwo\cos2(\dphi_1-2\dphi_2)+2\vtrig\vtwo\vvone\cos2(2\dphi_1-\dphi_2)+\\
2\vone\vtwo\vvtrig\cos2(\dphi_1+\dphi_2)-\\
4(\vtrig)^2\vone\vtwo\cos(2\dphi_1)\cos(2\dphi_2)-4\vtrig\vvtrig\vvone\vtwo\cos(4\dphi_1)\cos(2\dphi_2)-\\
4\vtrig\vvtrig\vone\vvtwo\cos(2\dphi_1)\cos(4\dphi_2)-4(\vvtrig)^2\vvone\vvtwo\cos(4\dphi_1)\cos(4\dphi_2)
\ea
\right).
\eea
If the true three-particle jet-correlation signal is given by Eq.~(\ref{eq:Jhat3factorize}), then that with the scaled background (by background normalization scheme) becomes
\bea
\Jhat_3(\dphi_1,\dphi_2)
&=&\left(\Jhat_2(\dphi_1)-\delta B_1\left[1+2\vtrig\vone\cos(2\dphi_1)+2\vvtrig\vvone\cos(4\dphi_1)\right]\right)\times\nonumber\\
&&\left(\Jhat_2(\dphi_2)-\delta B_1\left[1+2\vtrig\vtwo\cos(2\dphi_2)+2\vvtrig\vvtwo\cos(4\dphi_2)\right]\right)\nonumber\\
&&-[2B_1\delta B_1+(\delta B_1)^2]\left(
\ba{l}
2\vone\vtwo\cos2(\dphi_1-\dphi_2)+2\vvone\vvtwo\cos4(\dphi_1-\dphi_2)+\\
2\vtrig\vone\vvtwo\cos2(\dphi_1-2\dphi_2)+2\vtrig\vtwo\vvone\cos2(2\dphi_1-\dphi_2)+\\
2\vone\vtwo\vvtrig\cos2(\dphi_1+\dphi_2)-\\
4(\vtrig)^2\vone\vtwo\cos(2\dphi_1)\cos(2\dphi_2)-4\vtrig\vvtrig\vvone\vtwo\cos(4\dphi_1)\cos(2\dphi_2)-\\
4\vtrig\vvtrig\vone\vvtwo\cos(2\dphi_1)\cos(4\dphi_2)-4(\vvtrig)^2\vvone\vvtwo\cos(4\dphi_1)\cos(4\dphi_2)
\ea
\right).
\label{eq:Jhat3_Bnorm}
\eea

We postpone the discussion of the above result to later. 
We now proceed to the calculation of the lab-frame three-particle cumulant. From Eqs.~(\ref{eq4a}), (\ref{eq7a}), (\ref{eq7b}), (\ref{eq12}), and using
\be
\rho_2(\dphi_1,\dphi_2)=\rho_1^2\left[1+2v_2^2\cos2(\dphi_1-\dphi_2)+2v_4^2\cos4(\dphi_1-\dphi_2)\right]=(\rho_1/B_1)^2B_2(\dphi_1-\dphi_2),\label{eq27}
\ee
we obtain
\be
\rhohat_3(\dphi_1,\dphi_2)=J_3(\dphi_1,\dphi_2)-\left[J_2(\dphi_1)+J_2(\dphi_2)\right]\left(B_1+\mean{\Jhat_2}\right)-\left(1+\mean{\Jhat_2}/B_1\right)^2B_2(\dphi_1-\dphi_2)+2\left(B_1+\mean{\Jhat_2}\right)^2.\label{eq28}
\ee
Taking the difference between Eqs.~(\ref{eq28}) and~(\ref{eq6b}), using Eqs.~(\ref{eq22}) and (\ref{eq25}), we have
\bea
\Delta&=&\rhohat_3(\dphi_1,\dphi_2)-\Jhat_3(\dphi_1,\dphi_2)\nonumber\\
&=&J_2(\dphi_1)\left(B_1\left[1+2\vtrig\vtwo\cos(2\dphi_2)+2\vvtrig\vvtwo\cos(4\dphi_2)\right]-B_1-\mean{\Jhat_2}\right)\nonumber\\
&&+J_2(\dphi_2)\left(B_1\left[1+2\vtrig\vone\cos(2\dphi_1)+2\vvtrig\vvone\cos(4\dphi_1)\right]-B_1-\mean{\Jhat_2}\right)\nonumber\\
&&+B_1^2\left[
\ba{ll}
1&+2\vtrig\vone\cos(2\dphi_1)+2\vtrig\vtwo\cos(2\dphi_2)+2\vone\vtwo\cos2(\dphi_1-\dphi_2)\\
 &+2\vvtrig\vvone\cos(4\dphi_1)+2\vvtrig\vvtwo\cos(4\dphi_2)+2\vvone\vvtwo\cos4(\dphi_1-\dphi_2)\\
 &+2\vtrig\vone\vvtwo\cos2(\dphi_1-2\dphi_2)+2\vtrig\vtwo\vvone\cos2(2\dphi_1-\dphi_2)+2\vone\vtwo\vvtrig\cos2(\dphi_1+\dphi_2)
\ea
\right]\nonumber\\
&&-\left(B_1^2+2B_1\mean{\Jhat_2}+\mean{\Jhat_2}^2\right)\left[1+2\vone\vtwo\cos2(\dphi_1-\dphi_2)+2\vvone\vvtwo\cos4(\dphi_1-\dphi_2)\right]\nonumber\\
&&+2\left(B_1^2+2B_1\mean{\Jhat_2}+\mean{\Jhat_2}^2\right)\nonumber\\
&&-2B_1^2\left[1+2\vtrig\vone\cos(2\dphi_1)+2\vvtrig\vvone\cos(4\dphi_1)\right]\left[1+2\vtrig\vtwo\cos(2\dphi_2)+2\vvtrig\vvtwo\cos(4\dphi_2)\right]\nonumber\\
&=&-\mean{\Jhat_2}\left(\Jhat_2(\dphi_1)+\Jhat_2(\dphi_2)-\mean{\Jhat_2}\right)\nonumber\\
&&+B_1\left(\Jhat_2(\dphi_1)-\mean{\Jhat_2}\right)\left[2\vtrig\vtwo\cos(2\dphi_2)+2\vvtrig\vvtwo\cos(4\dphi_2)\right]\nonumber\\
&&+B_1\left(\Jhat_2(\dphi_2)-\mean{\Jhat_2}\right)\left[2\vtrig\vone\cos(2\dphi_1)+2\vvtrig\vvone\cos(4\dphi_1)\right]\nonumber\\
&&-\mean{\Jhat_2}\left(2B_1+\mean{\Jhat_2}\right)\left[2\vone\vtwo\cos2(\dphi_1-\dphi_2)+2\vvone\vvtwo\cos4(\dphi_1-\dphi_2)\right]\nonumber\\
&&+B_1^2\left[2\vtrig\vone\vvtwo\cos2(\dphi_1-2\dphi_2)+2\vtrig\vtwo\vvone\cos2(2\dphi_1-\dphi_2)+2\vone\vtwo\vvtrig\cos2(\dphi_1+\dphi_2)\right].
\label{eq30}
\eea
Assuming factorization of the three-particle correlation signal in Eq.~(\ref{eq:jetGaus}), the three-particle cumulant is therefore
\be\ba{rl}
\rhohat_3(\dphi_1,\dphi_2)=&\left[\Jhat_2(\dphi_1)-\mean{\Jhat_2}\right]\left[\Jhat_2(\dphi_2)-\mean{\Jhat_2}\right]\\
&+B_1\left(\Jhat_2(\dphi_1)-\mean{\Jhat_2}\right)\left[2\vtrig\vtwo\cos(2\dphi_2)+2\vvtrig\vvtwo\cos(4\dphi_2)\right]\\
&+B_1\left(\Jhat_2(\dphi_2)-\mean{\Jhat_2}\right)\left[2\vtrig\vone\cos(2\dphi_1)+2\vvtrig\vvone\cos(4\dphi_1)\right]\\
&-\mean{\Jhat_2}\left(2B_1+\mean{\Jhat_2}\right)\left[2\vone\vtwo\cos2(\dphi_1-\dphi_2)+2\vvone\vvtwo\cos4(\dphi_1-\dphi_2)\right]\\
&+B_1^2\left[2\vtrig\vone\vvtwo\cos2(\dphi_1-2\dphi_2)+2\vtrig\vtwo\vvone\cos2(2\dphi_1-\dphi_2)+2\vone\vtwo\vvtrig\cos2(\dphi_1+\dphi_2)\right].
\ea\label{eq:rho3hat_flow}\ee
\ew
The first line in the r.h.s. of Eq.~(\ref{eq:rho3hat_flow}) is the three-particle cumulant result for the case of no anisotropic flow correlation, and the fifth line is the three-particle cumulant due to irreducible anisotropic flow correlation [as in Eq.~(\ref{eq26})]. However, there are additional terms in the rest of Eq.~(\ref{eq:rho3hat_flow}) r.h.s., which arise from the coupling between jet-correlation and anisotropic flow.

We are now ready to compare the jet-correlation method and the lab-frame cumulant method for the realistic case of jet-correlation together with anisotropic flow background. We again use the jet-model of case (B) in Eq.~(\ref{eq:caseB}), but with additional anisotropic flow of magnitudes given by Eq.~(\ref{eq:flow_value}). Namely,
\be\left\{
\ba{l}
N_1=0.7,N_2=1.2,\sigma_1=0.4,\sigma_2=0.7,\theta=1;\\
B_1=\Btrue=150/2\pi, \delta B_1=0.12;\\
\vtrig=7.5\%,\vone=\vtwo=5\%,v_4=v_2^2.
\ea
\right.\label{eq32}\ee

\begin{figure*}[hbt]
\centerline{
\psfig{file=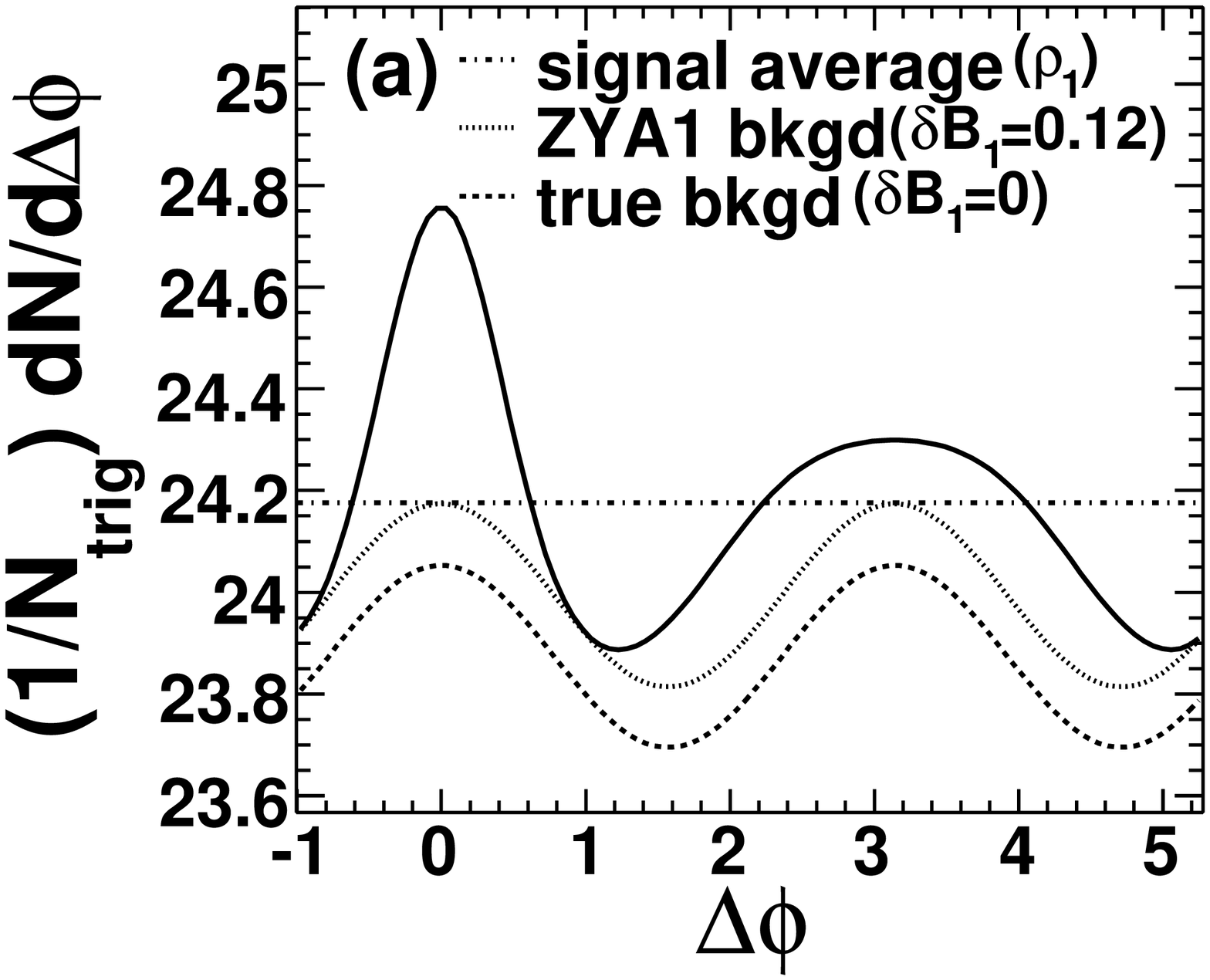,width=0.25\textwidth}
\psfig{file=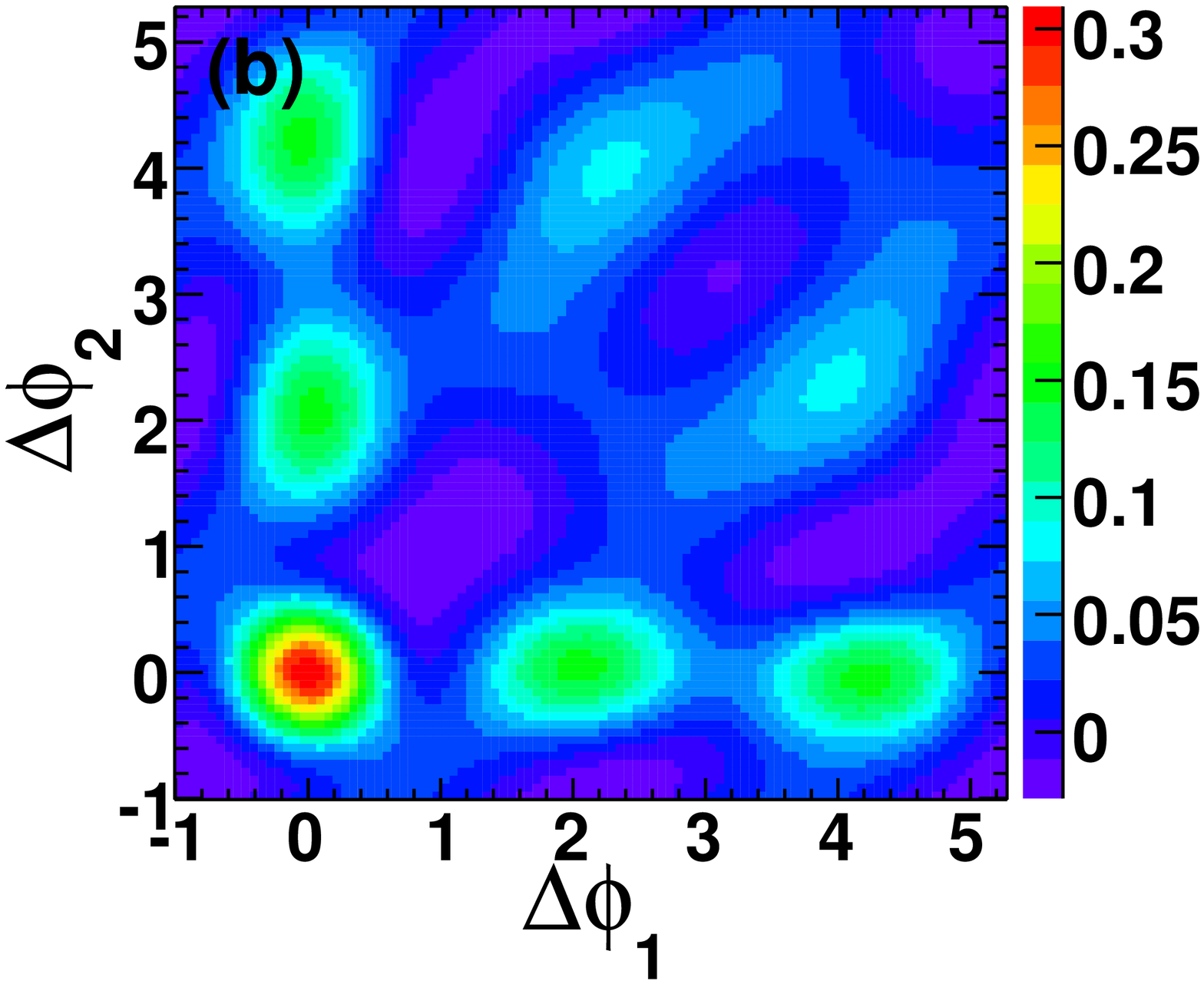,width=0.25\textwidth}
\psfig{file=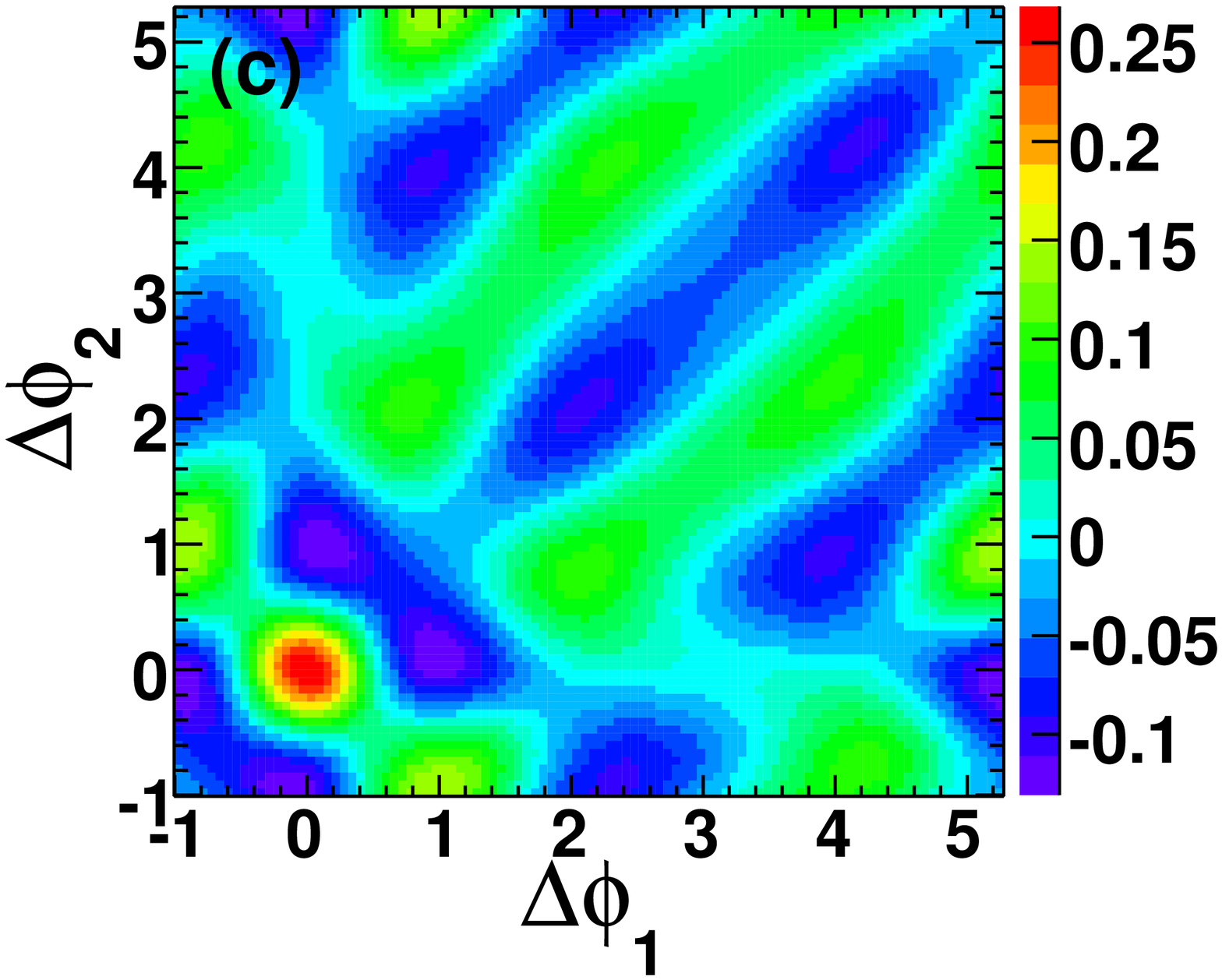,width=0.25\textwidth}
\psfig{file=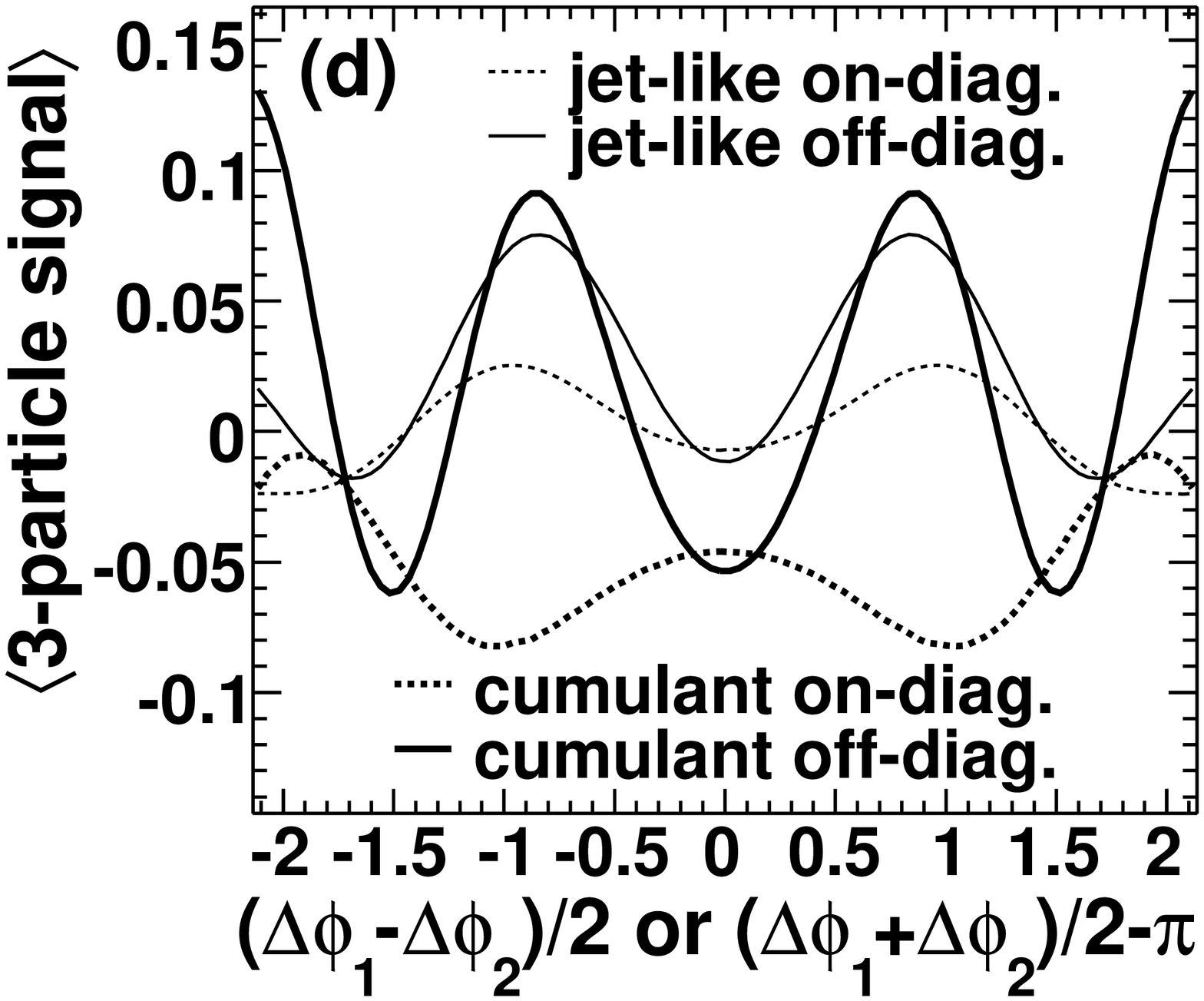,width=0.25\textwidth}
}
\caption{(Color online) Jet-correlation model with realistic jets and Mach-cones together with anisotropic flow background, Eq.~(\ref{eq32}). (a) Two-particle correlation signal, $J_2(\dphi)$ by Eq.~(\ref{eq:Jhat2_flow}) (solid curve) atop a flow modulated background, $B_2(\dphi)$ by Eq.~(\ref{eq:B2_flow}) (dashed curve). The dotted curve is the (over)estimated background (by $\delta B_1=0.12$) to match the signal at the minimum (ZYA1). The dash-dotted line is the average of the raw correlation signal, $\rho_1$. (b) The three-particle jet-correlation after subtraction of the ZYA1-normalized background [represented by the dotted curve in panel (a)]. (c) The lab-frame three-particle cumulant. (d) Away-side on- and off-diagonal projections of the jet-like result in panel (b) in thin curves and cumulant result in panel (c) in thick curves. The color bars on the right in panels (b) and (c) are three-particle correlation magnitudes.}
\label{fig5}
\end{figure*}

\begin{figure*}[hbt]
\centerline{
\psfig{file=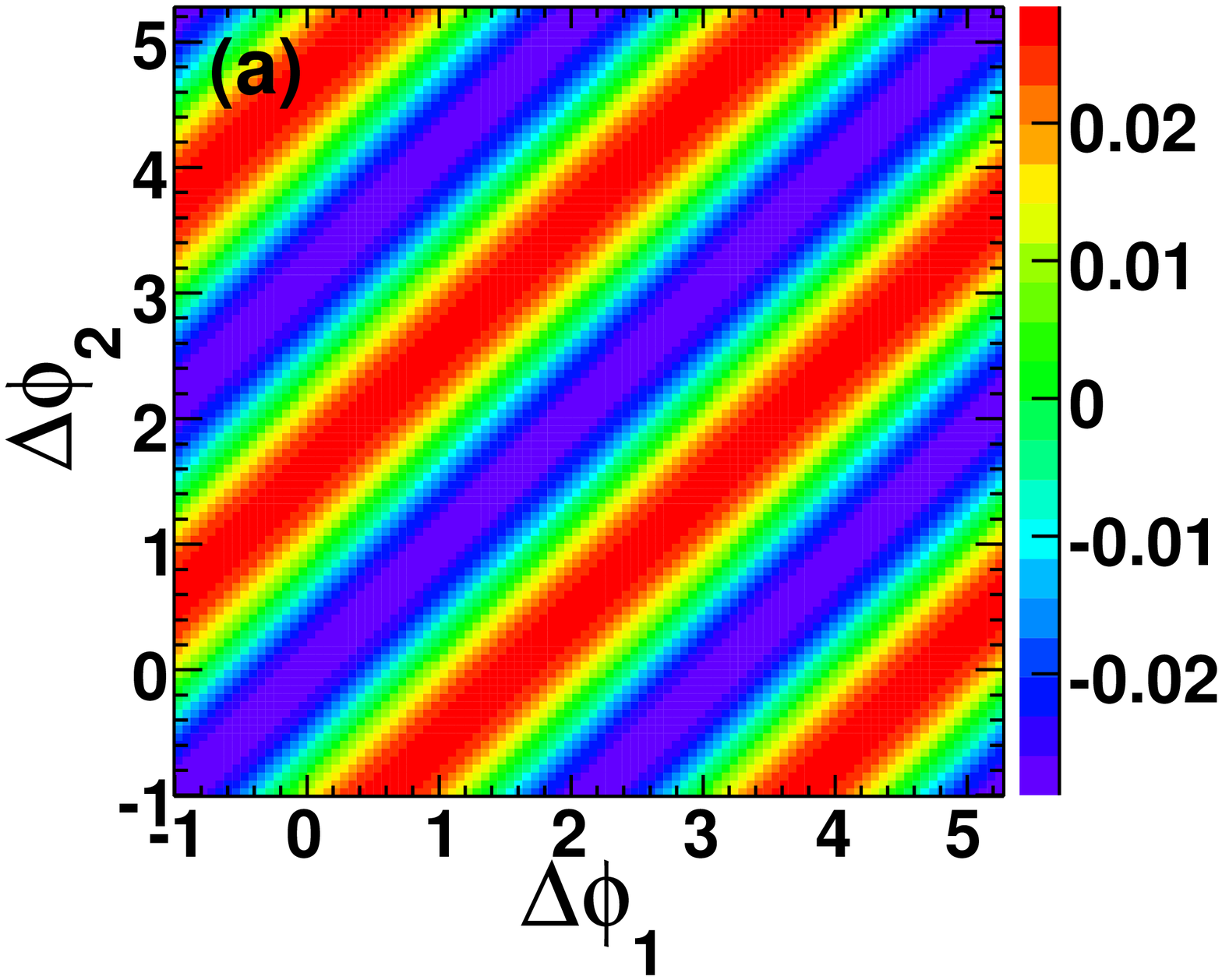,width=0.25\textwidth}
\psfig{file=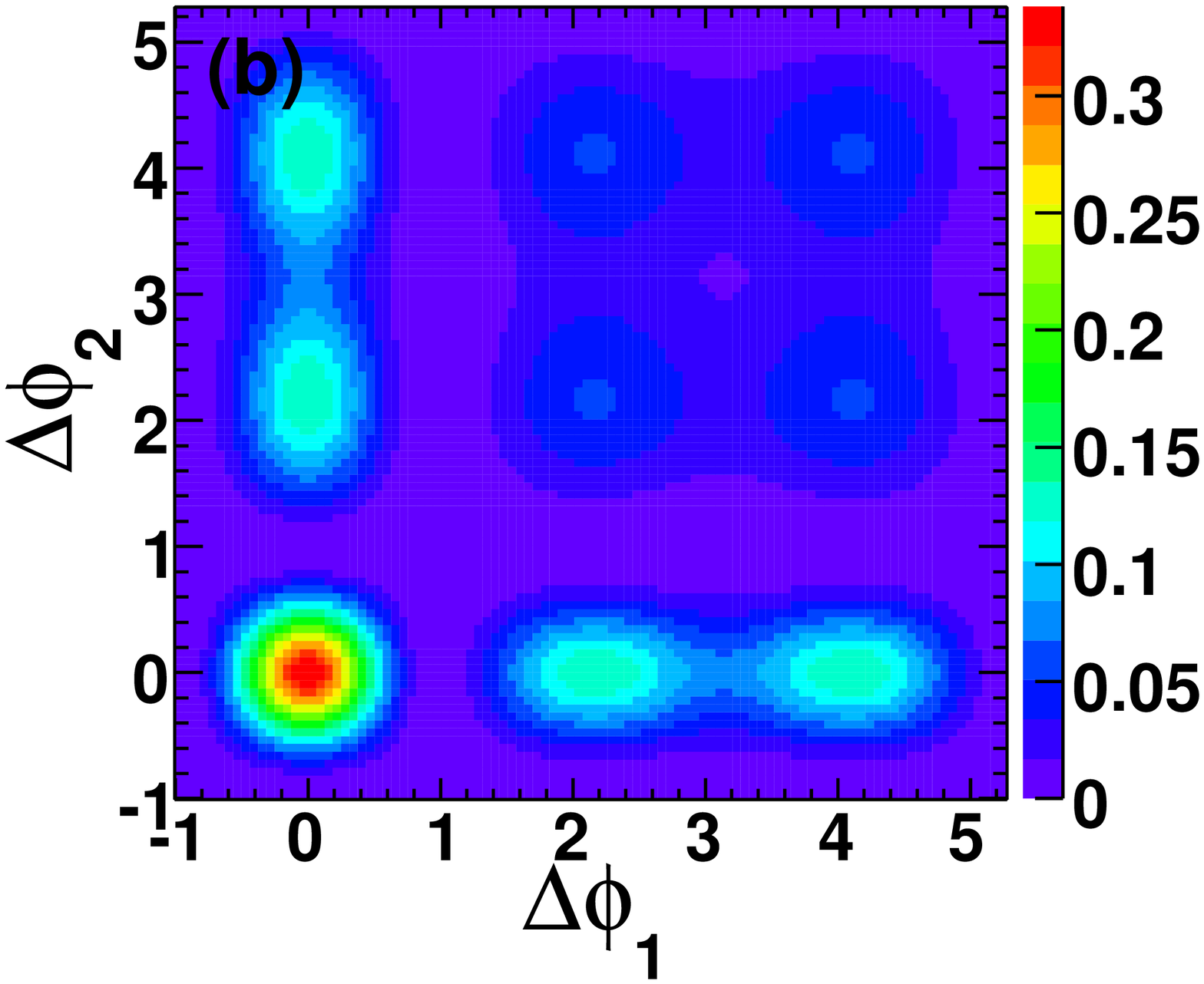,width=0.25\textwidth}
\psfig{file=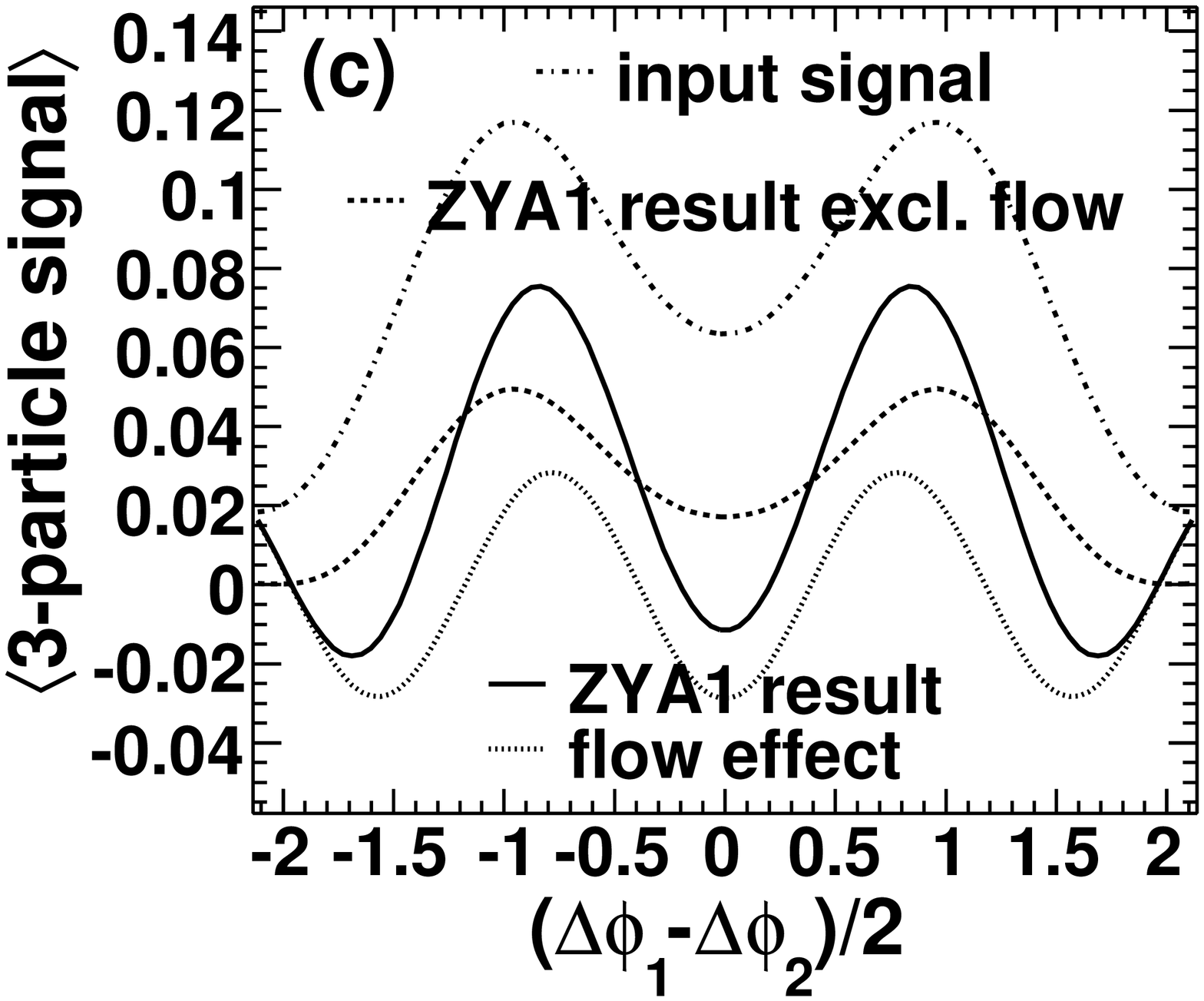,width=0.25\textwidth}
\psfig{file=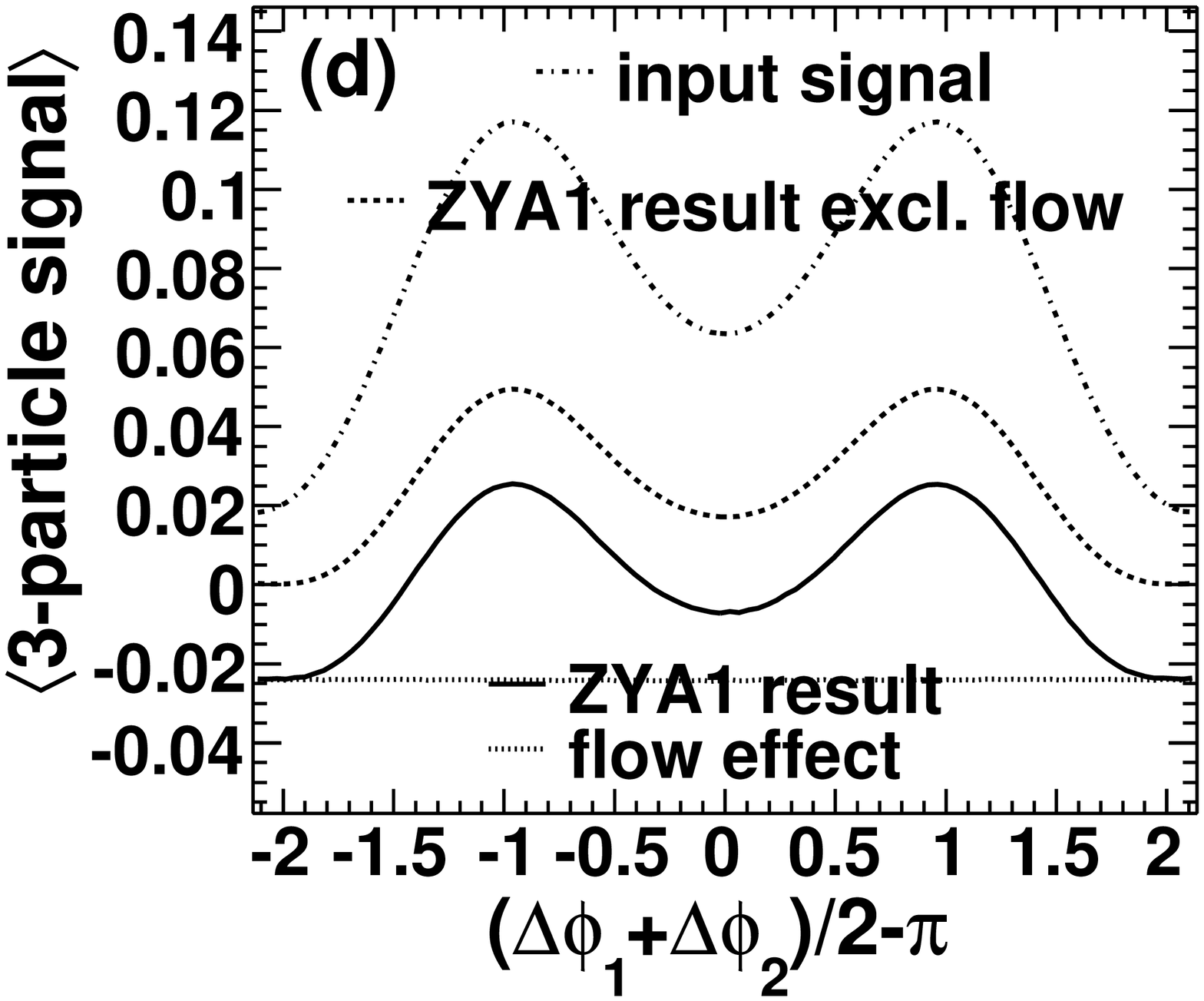,width=0.25\textwidth}
}
\caption{(Color online) Jet-correlation method details using jet-correlation model with realistic jets and Mach-cones together with anisotropic flow background, Eq.~(\ref{eq32}). (a) The error introduced by an inaccurate background level, the second term in r.h.s.~of Eq.~(\ref{eq:Jhat3_Bnorm}). (b) Illustration of what the constructed three-particle correlation would be if the over-subtracted jet-correlated particles have the same anisotropic flow as the medium background particles so that the inaccurate background would not introduce structures shown in the left panel. The three-particle jet-correlation would be similar to the input in shape with a reduced magnitude, and would be similar to that constructed in the uniform background example shown in Fig.~\ref{fig2}(b). (c),(d) Away-side off-diagonal (c) and on-diagonal (d) projections of the input Mach-cone signal [i.e., as same as that in Fig.~\ref{fig2}(b)] in dash-dotted curves, of the structure of the flow effect in panel (a) in dotted curves introduced by an over-subtracted background level, of the three-particle signal in panel (b) in dashed curves that would be constructed if no structure was introduced by the inaccurate background level, and in solid curves of the obtained final three-particle signal using ZYA1 background subtraction in Fig.~\ref{fig5}(b). The solid curves in (c) and (d) are as same as the thin dashed and solid curve in Fig.~\ref{fig5}(d), respectively. The color bars on the right in panels (a) and (b) are three-particle correlation magnitudes.}
\label{fig6}
\end{figure*}

We first examine the effect of an imperfect background estimation (by ZYA1 or ZYAM normalization) on the final three-particle jet-correlation signal given by Eq.~(\ref{eq:Jhat3_Bnorm}). Figure~\ref{fig5}(a) shows in the solid curve the raw two-particle correlation, $J_2(\dphi)$ given by Eqs.~(\ref{eq:Jhat2_flow}) and (\ref{eq:jetGaus}) with parameters in Eq.~(\ref{eq32}), and in the dashed curve the flow modulated background of Eq.~(\ref{eq:B2_flow}). The dotted curve shows the scaled background by ZYA1. Figure~\ref{fig5}(b) shows the final three-particle jet-like correlation obtained by Eq.~(\ref{eq:Jhat3_Bnorm}) after subtraction of the scaled backgrounds. The result is significantly distorted from the input correlation [as shown in Fig.~\ref{fig2}(b)], however, the Mach-cone structure seems still evident. The distortion is due to flow effect caused by subtraction of an incorrect background level by ZYA1. Figure~\ref{fig5}(d) shows the away-side on- and off-diagonal projections in thin curves. The on- and off-diagonal projections are not the same any more; the symmetry in the input Mach-cone signal is lost. The flow effect due to over-subtraction of the background level is different between on- and off-diagonal projection; the shape of the on-diagonal projection is not affected but that of the off-diagonal project is (also see below). The net effect is that the peaks in the off-diagonal projection are pulled toward the flow peaks at $\pm\pi/4$. The effect of flow due to incorrect subtraction of the background level will be further discussed below.

Figure~\ref{fig5}(c) shows the lab-frame three-particle cumulant by Eq.~(\ref{eq:rho3hat_flow}). The cumulant result is very different from the three-particle jet-correlation result shown in Fig.~\ref{fig2}(b) or Fig.~\ref{fig5}(b). The structure of the cumulant is complex. Figure~\ref{fig5}(d) also shows the away-side on-diagonal and off-diagonal projections of the cumulant result in thick curves. The on- and off-diagonal projections are out of phase; the on-diagonal projection is even all negative. The input Mach-cone signal, which consists of two off-center peaks in both on- and off-diagonal directions, is not observable.

The distortion of the constructed jet-like three-particle correlation is due to the overestimated background level by $\delta B_1/B_1\sim0.5\%$. The jet-correlation signal from Eq.~(\ref{eq:jetGaus}) does not contain anisotropic flow. However, part of the signal is now included into the background which is all taken to be flow-modulated. This introduces flow structures in the over-subtracted part of the three-particle background which should be really part of the signal that is uniform. The reverse of the over-subtracted background is given by the last line of r.h.s.~of Eq.~(\ref{eq:Jhat3_Bnorm}); we show this in Fig.~\ref{fig6}(a). 

If the number of jet-correlated particles also varies with reaction plane in the same way as the medium particles, then the constructed three-particle jet-correlation would be similar to the input correlation except an approximately constant reduction in the signal magnitude. In other words, the final signal would be as same as given by the first line of r.h.s.~of Eq.~(\ref{eq:Jhat3_Bnorm}). This is shown in Fig.~\ref{fig6}(b). The signal shape is not significantly altered from the true signal in Fig.~\ref{fig2}(b). The argument can be turned around: if the medium particle background is uniform relative to reaction plane, then the over-subtracted background due to ZYA1 normalization will not introduce structures and the final three-particle correlation would be similar to the true one except a constant reduction in the signal magnitude [Fig.~\ref{fig3}(a)]. The results in Figs.~\ref{fig3}(a) and~\ref{fig6}(b) are indeed very similar. 

Figures~\ref{fig6}(c) and (d) show the effect of flow due to subtraction of an incorrect background level in off- and on-diagonal projection, respectively. The dash-dotted curve shows the input Mach-cone signal [projection of Fig.~\ref{fig2}(b)]; the dashed curve shows the projection of the jet-like three-particle correlation by ZYA1 background subtraction but excluding flow effect [projection of Fig.~\ref{fig6}(b)]; the dotted curve shows the projection of the flow effect in Fig~\ref{fig6}(a); and the thick solid curve shows the projection of the final result [projection of Fig.~\ref{fig5}(b)], which is the sum of the dashed and dotted curves. The ZYA1 background subtraction causes an reduction in the three-particle correlation magnitude. The flow effect changes the magnitude of the on-diagonal projection, but does not affect its shape. However, the flow mismatch between jet-correlated particles (no flow in our particular model example) and medium background particles (finite flow in our model example) affects off-diagonal projection in both its magnitude and shape. The flow effect pulls the off-diagonal peak positions from the input $\pm1$ toward the flow peaks at $\pm\pi/4$.

\begin{figure*}[hbt]
\centerline{
\psfig{file=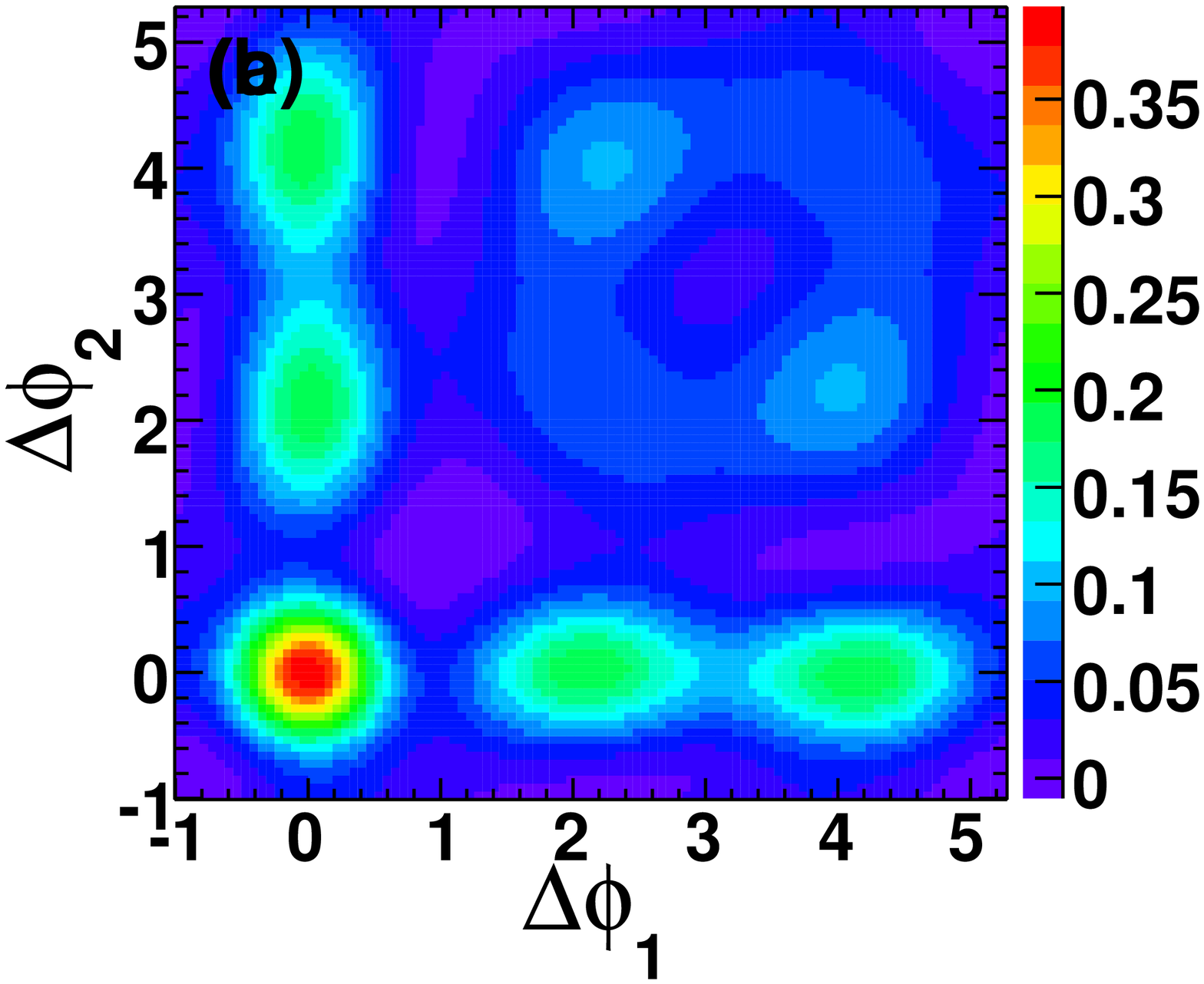,width=0.25\textwidth}
\psfig{file=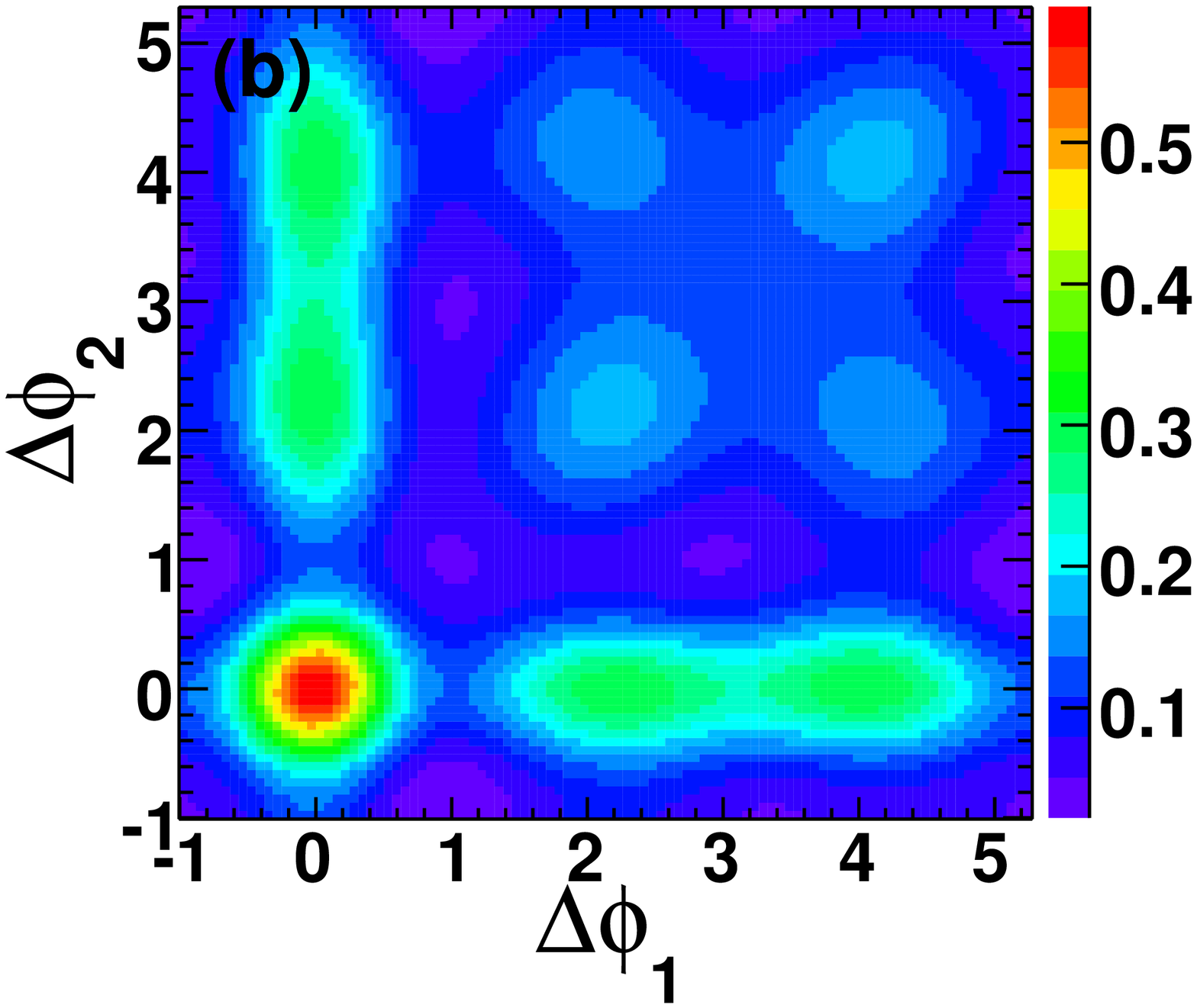,width=0.25\textwidth}
\psfig{file=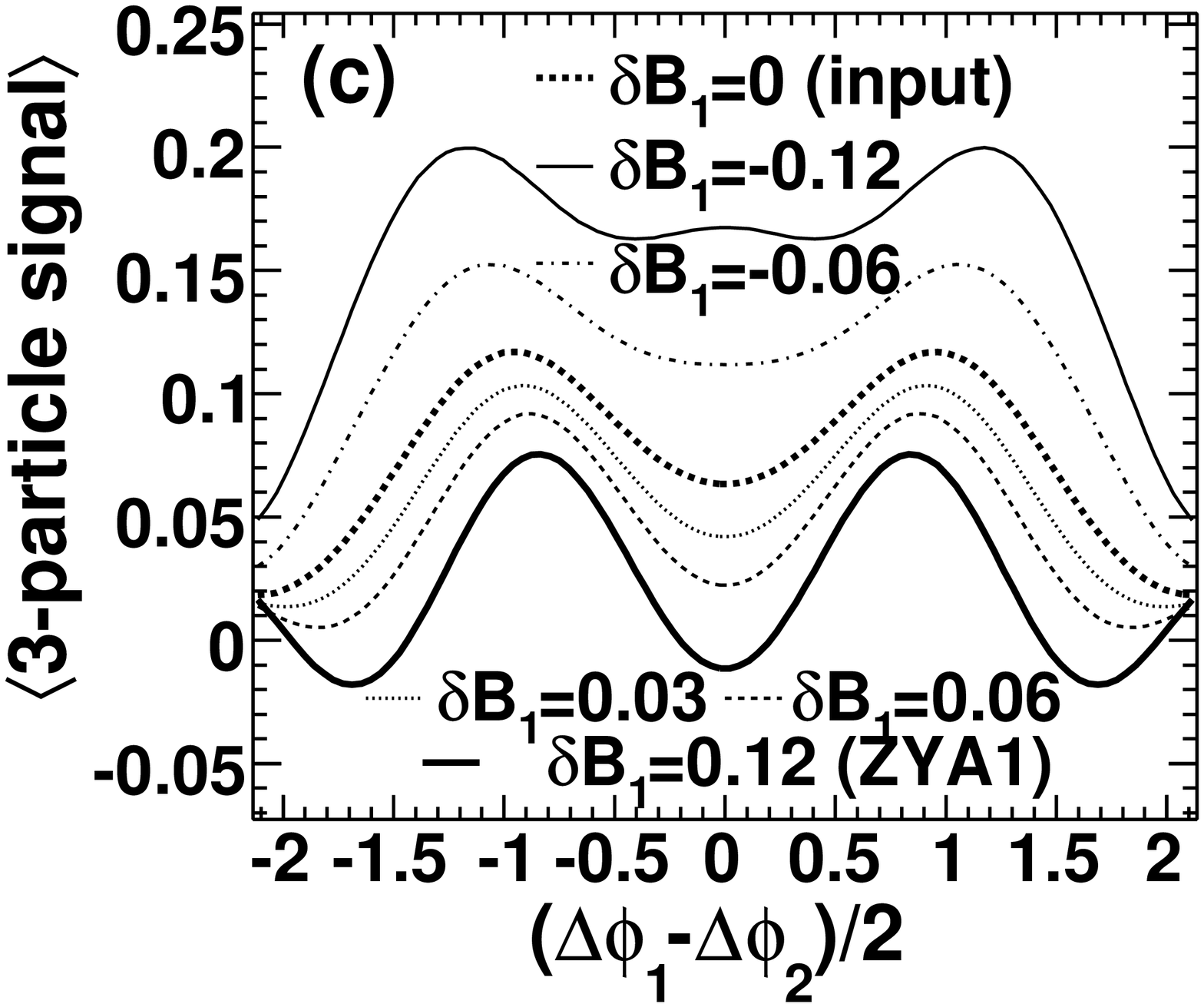,width=0.25\textwidth}
\psfig{file=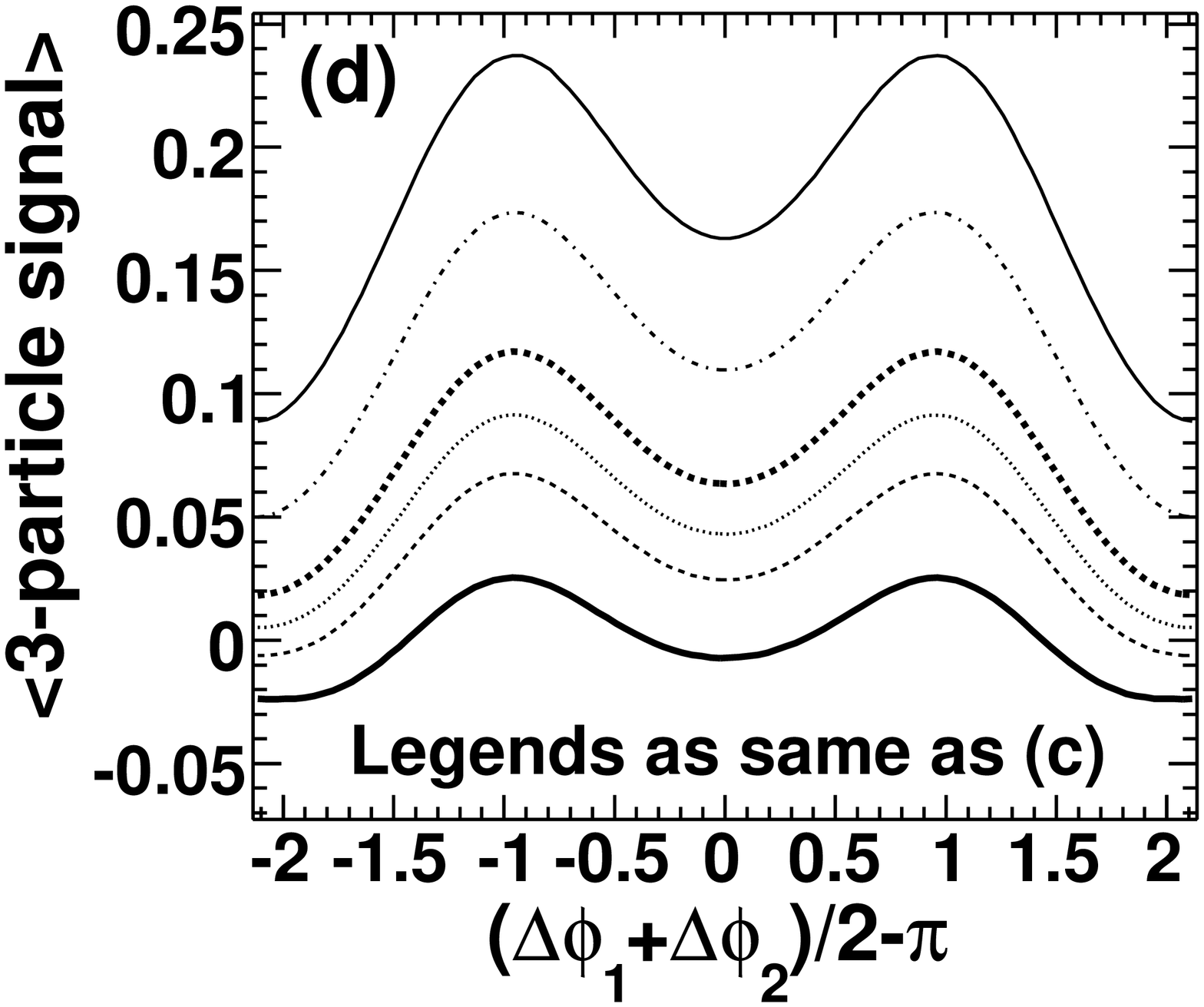,width=0.25\textwidth}
}
\caption{(Color online) Jet-correlation method background normalization effect using jet-correlation model with realistic jets and Mach-cones together with anisotropic flow background, Eq.~(\ref{eq32}). (a) The three-particle jet-like correlation signal after over-subtraction of a background half way to ZYA1 (i.e., by $\delta B_1=0.06$). (b) The three-particle jet-like correlation signal after under-subtraction of background by $\delta B_1=-0.06$). (c),(d) Away-side off-diagonal (c) and on-diagonal (d) projections of the jet-like correlation signals using different background normalizations by $\delta B_1$. The thick dashed curves ($\delta B_1=0$) are the input signal [and the same as the dashed curve in Fig.~\ref{fig2}(d)], and the thick solid curves ($\delta B_1=0.12$) are the final signal by ZYA1 normalization [and the same as the respective solid curves in Fig.~\ref{fig6}(c,d)]. The color bars on the right in panels (a) and (b) are three-particle correlation magnitudes.}
\label{fig7}
\end{figure*}

\begin{figure*}[hbt]
\centerline{
\psfig{file=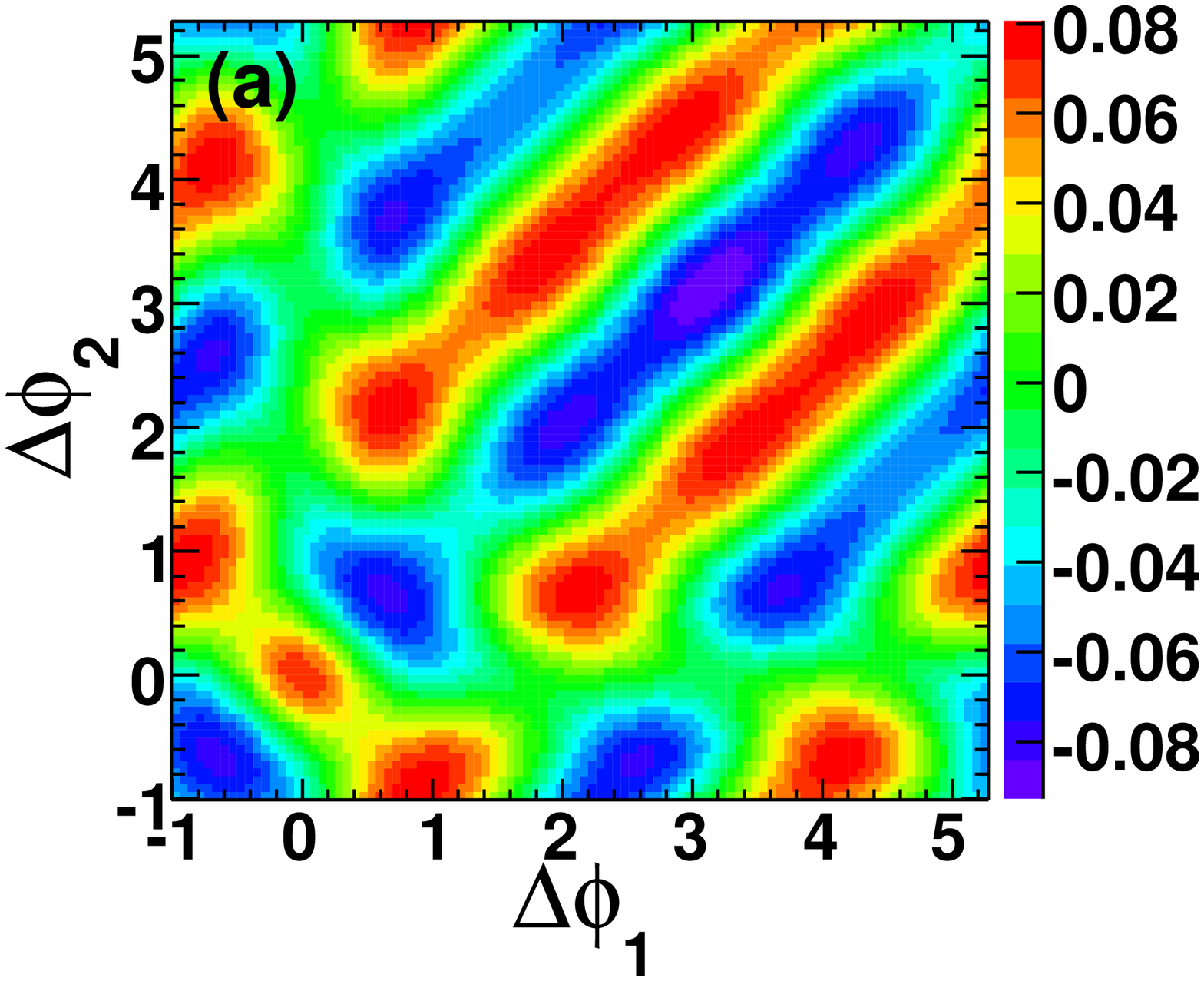,width=0.25\textwidth}
\psfig{file=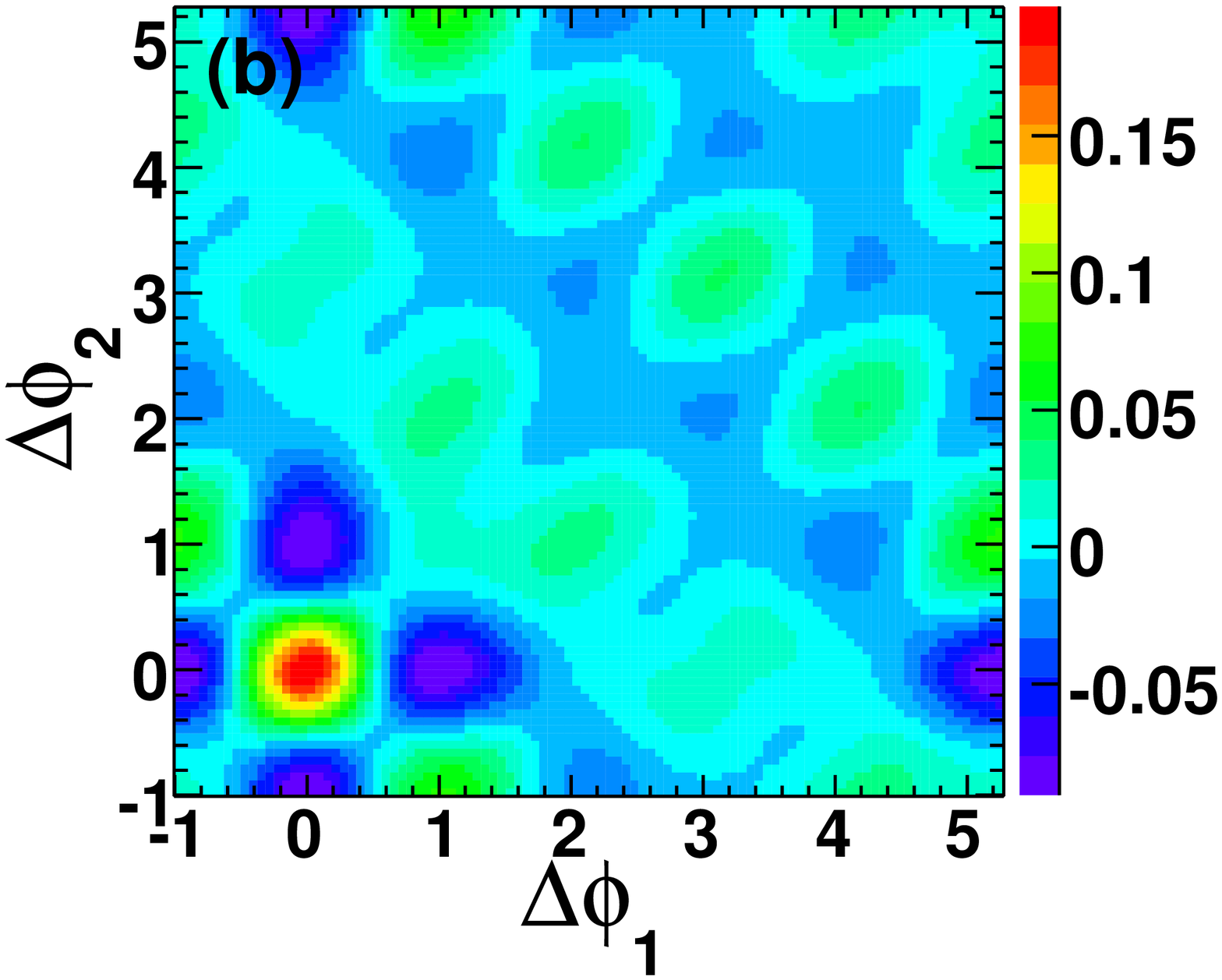,width=0.25\textwidth}
\psfig{file=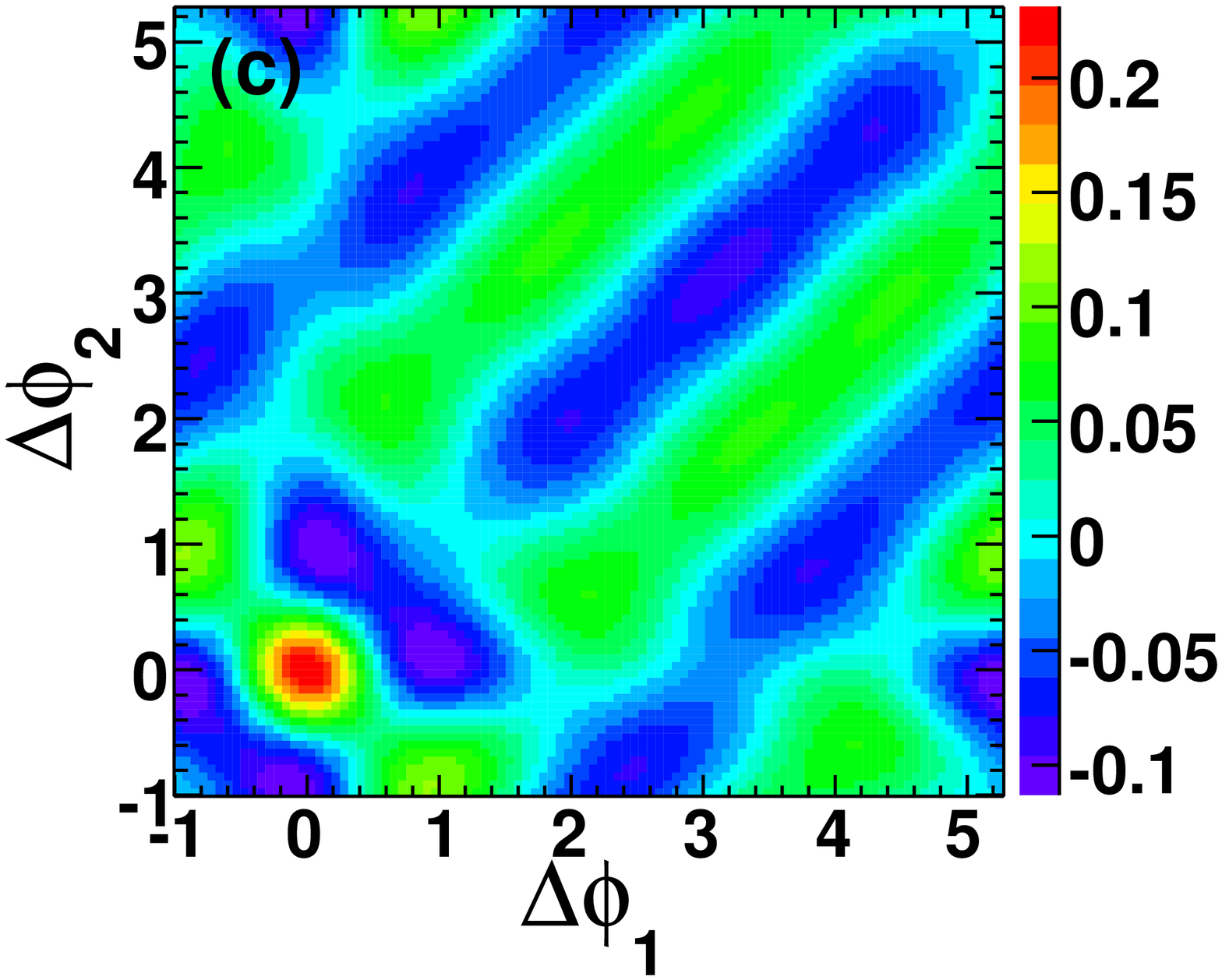,width=0.25\textwidth}
\psfig{file=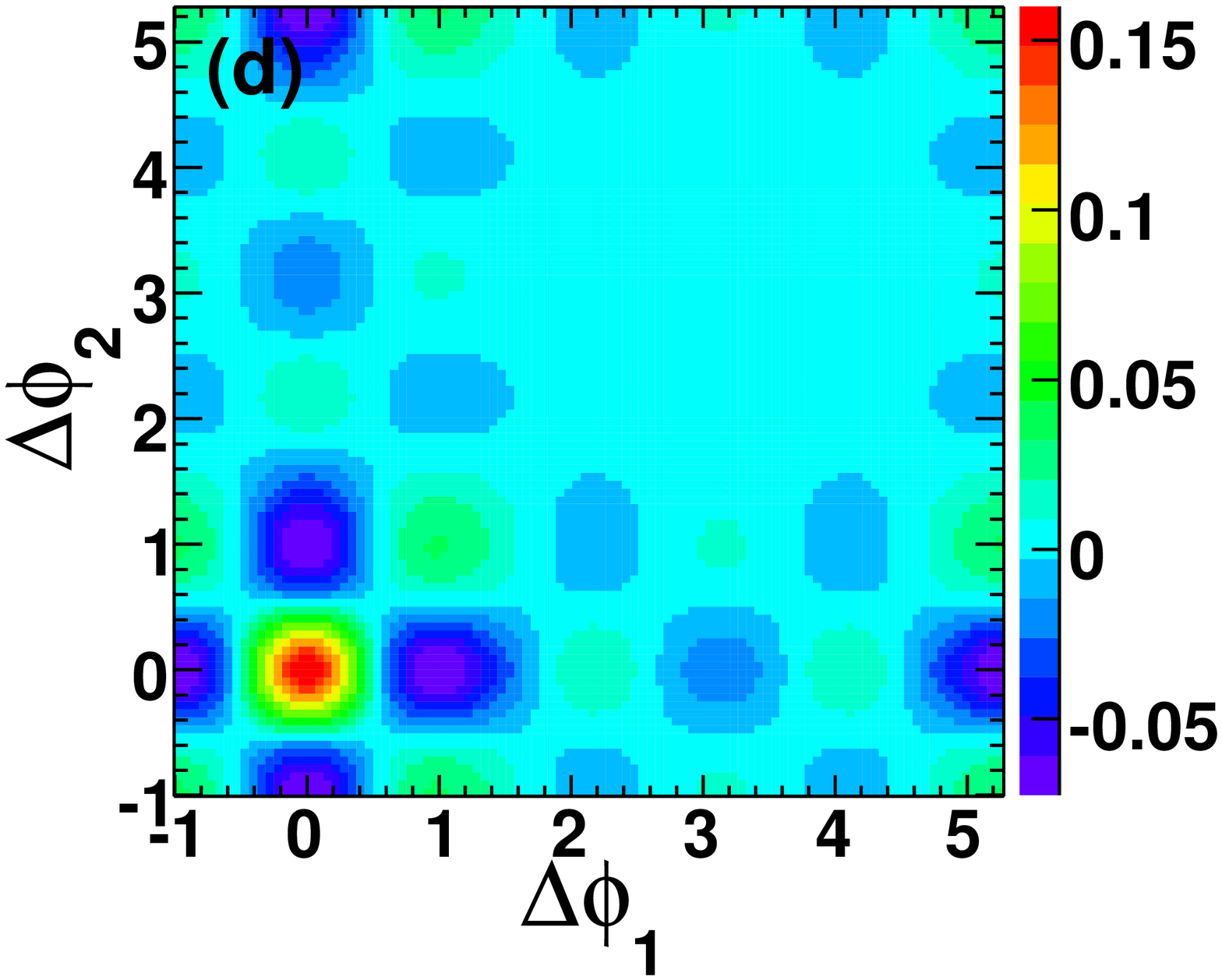,width=0.25\textwidth}
}
\centerline{
\psfig{file=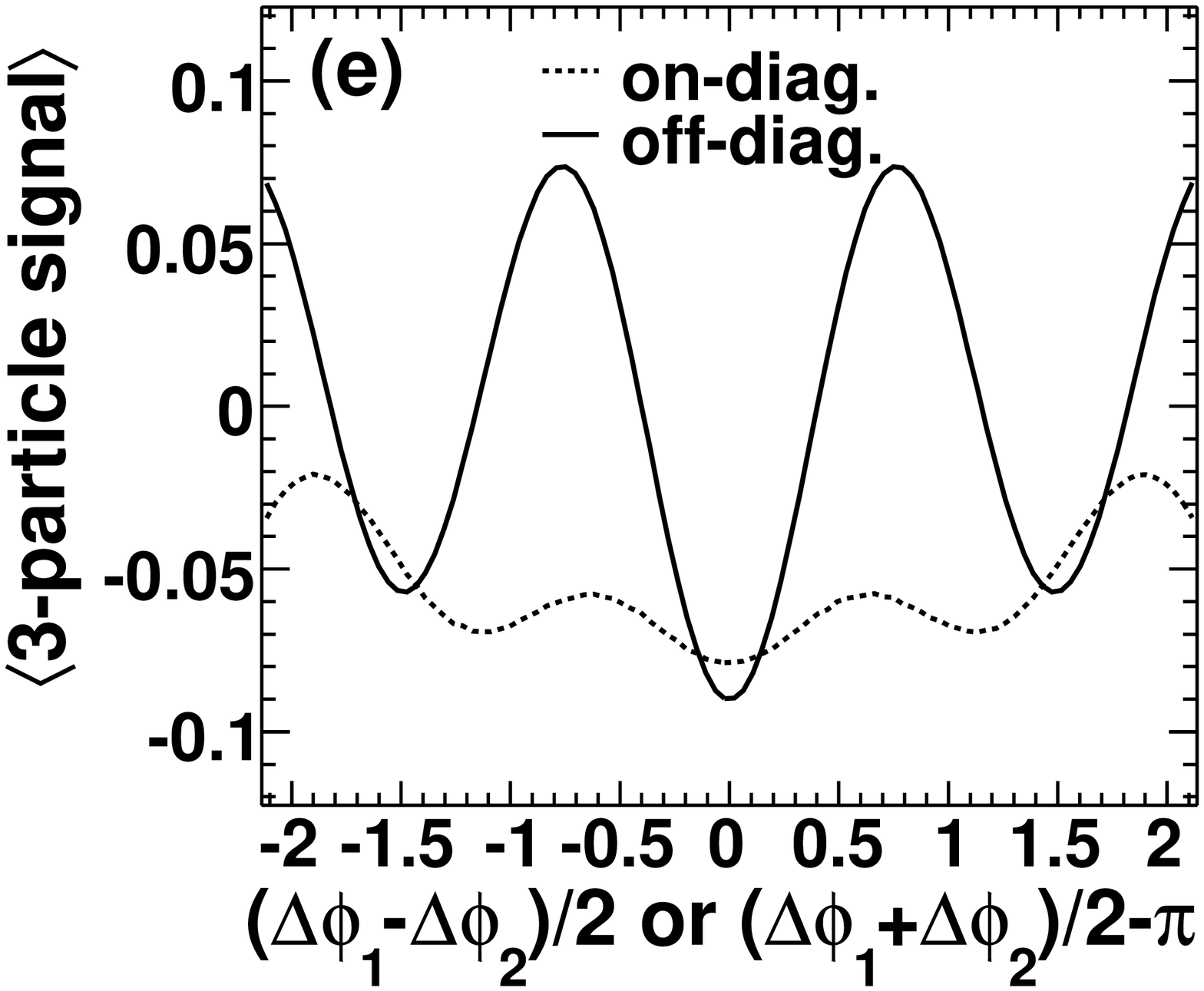,width=0.25\textwidth}
\psfig{file=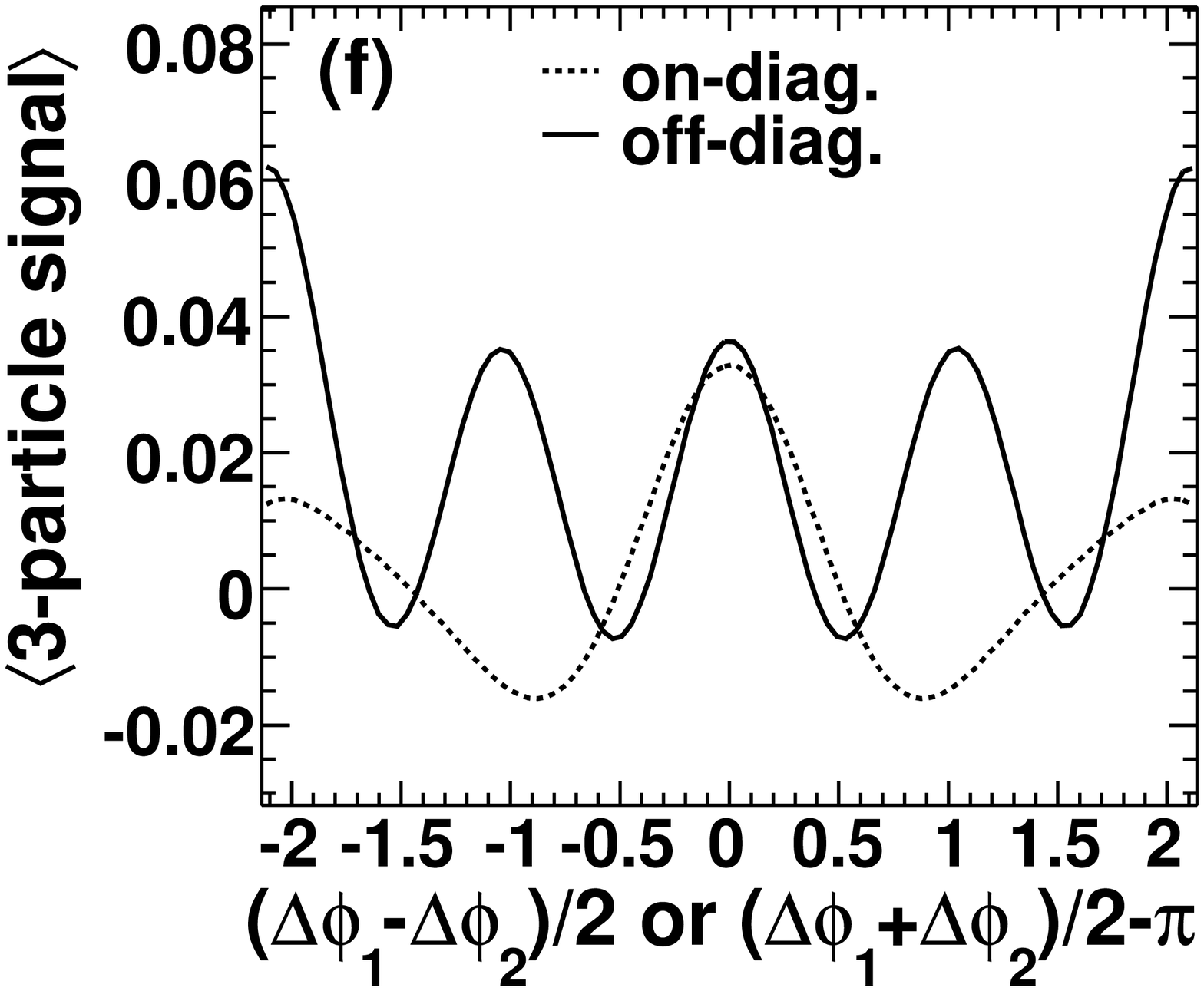,width=0.25\textwidth}
\psfig{file=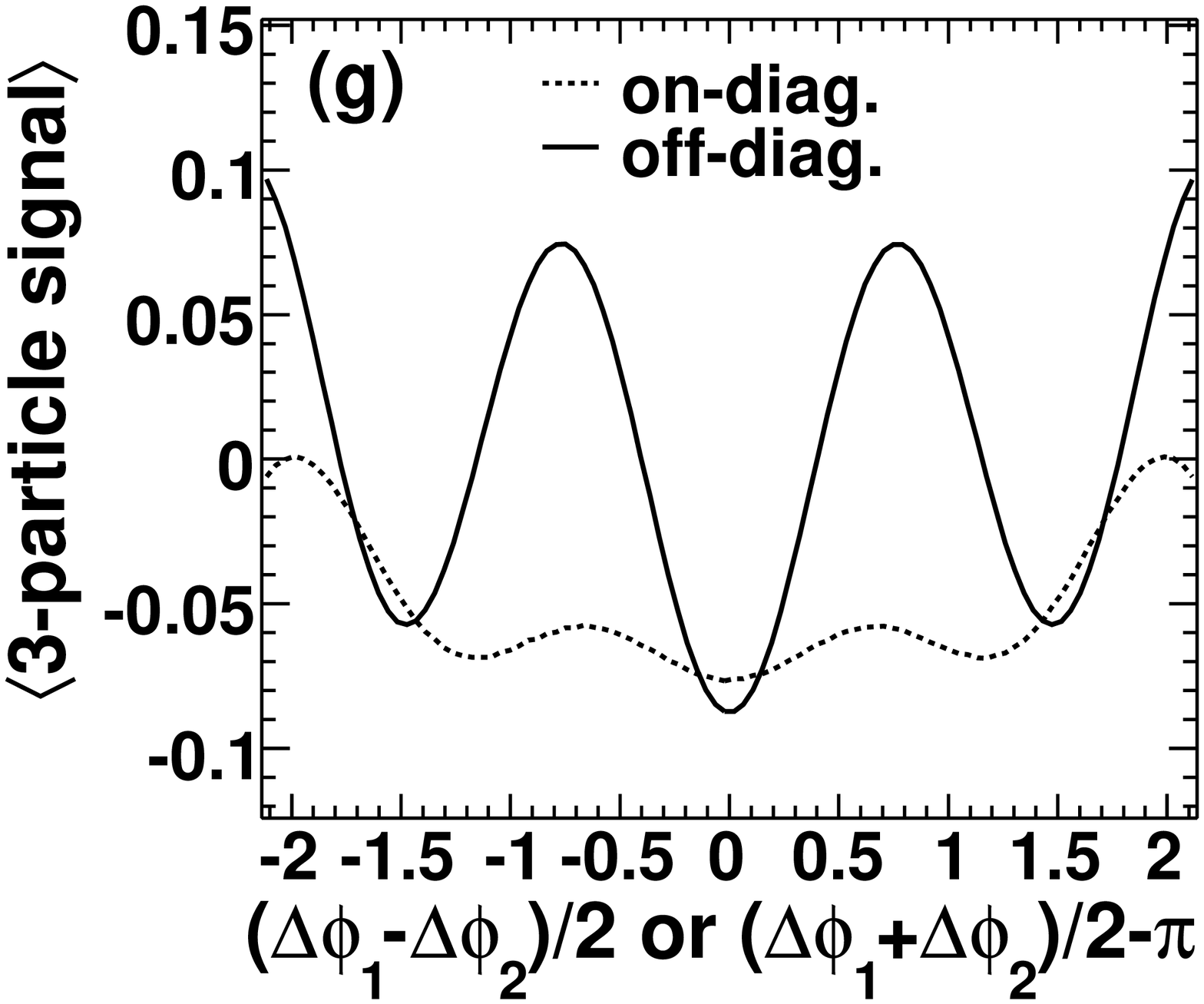,width=0.25\textwidth}
\psfig{file=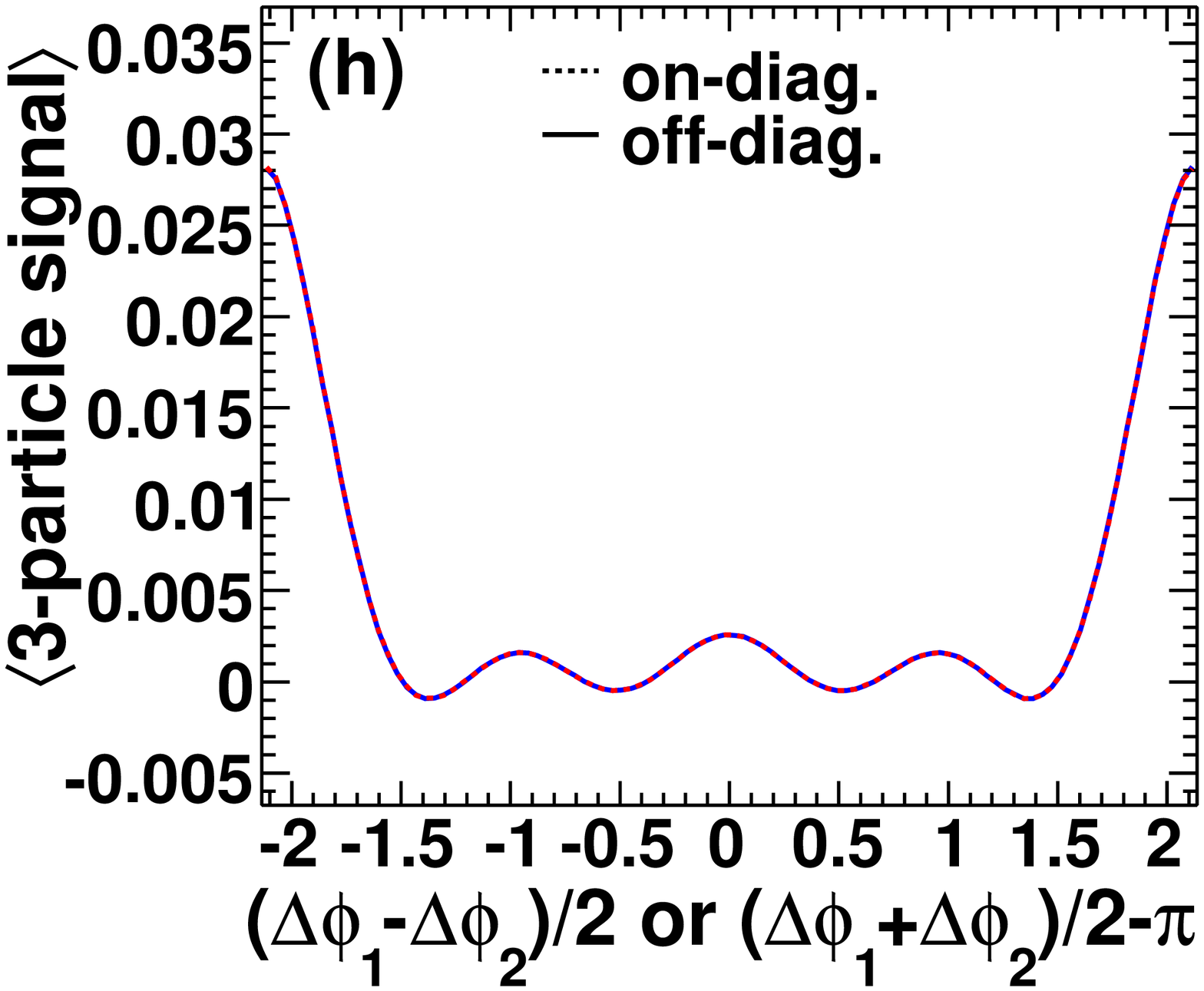,width=0.25\textwidth}
}
\caption{(Color online) Cumulant method details using jet-correlation model with realistic jets and Mach-cones together with anisotropic flow background, Eq.~(\ref{eq32}). (a) The jet-flow cross-terms in lab-frame three-particle cumulant [i.e., sum of lines 2, 3, and 4 of the r.h.s.~of Eq.~(\ref{eq:rho3hat_flow})]. (b) The lab-frame three-particle cumulant excluding the jet-flow cross terms [i.e., sum of lines 1 and 5 of the r.h.s.~of Eq.~(\ref{eq:rho3hat_flow})]. (c) The three-particle cumulant excluding the irreducible pure flow correlation terms [i.e., sum of lines 1, 2, 3, and 4 of the r.h.s.~of Eq.~(\ref{eq:rho3hat_flow})]. (d) The lab-frame three-particle cumulant excluding both the jet-flow cross-terms and the irreducible flow terms [i.e., line 1 of the r.h.s.~of Eq.~(\ref{eq:rho3hat_flow})]. This is similar to that constructed in the uniform background example shown in Fig.~\ref{fig2}(c). (e)-(h) Away-side on- and off-diagonal projections of panels (a)-(d), respectively. The projections are identical in (h). The color bars on the right in panels (a), (b), (c) and (d) are three-particle correlation magnitudes.}
\label{fig8}
\end{figure*}

In reality, the situation may be somewhere in-between: the jet-particles do have some variation relative to reaction plane, and the variation may not be as strong as that of the medium particles. If this is true, then the constructed three-particle jet-correlation signal should be somewhere in-between those in Figs.~\ref{fig5}(b) and~\ref{fig6}(b). We try to illustrate this by subtracting half of the jet-correlation pedestal as background (i.e., setting $\delta B_1=0.06$), effectively modulating the jet-correlation pedestal with half of the medium background flow. The result is shown in Fig.~\ref{fig7}(a). The Mach-cone signal is less distorted than in Fig.~\ref{fig5}(b), and the cone angle is closer to the input $\theta=1$. However, if jet-particles are more strongly modulated than background particles (due, for instance, to the recently observed ridge particles~\cite{Feng}), then the over-subtracted flow modulation by ZYA1 normalization would be too weak. The resulting jet-correlation shape may be mimicked in our jet-correlation model by subtracting a lower level of background (i.e., $\delta B_1<0$). We illustrate this in Fig.~\ref{fig7}(b) using $\delta B_1=-0.06$. As expected, the off-diagonal peaks are pushed away from the input angles $\pm1$ toward $\pm\pi/2$ (instead of $\pm\pi/4$ for $\delta B_1>0$), while the on-diagonal peak positions remain intact. On the other hand, it is also possible that the jet-particle modulation with respect to reaction plane is opposite to that of background particles, i.e., jet-particles have a negative elliptic flow parameter. This can come about at relatively low $\pt$ due to the path-length effect of jet-quenching; more low $\pt$ particles are generated in the direction perpendicular to reaction plane where more high $\pt$ particles are quenched. Such a scenario would make the flow effect due to ZYA1 background normalization more significant than shown in Fig.~\ref{fig5}(b), resulting in off-diagonal peaks closer to the flow peaks at $\pm\pi/4$.

If jet-correlation signal does not vary with reaction plane angle (as in our jet-correlation model), then the evolution from Fig.~\ref{fig7}(b) ($\delta B_1=-0.06$) to Fig.~\ref{fig2}(b) ($\delta B_1=0$) to Fig.~\ref{fig7}(a) ($\delta B_1=0.06$) to Fig.~\ref{fig5}(b) ($\delta B_1=0.12$) gives a good indication of the background normalization effect. The magnitude drops because of the larger background subtraction, and the shape evolves because of the flow effect due to subtraction of incorrect background. This evolution is better recapitulated in Figs.~\ref{fig7}(c) and (d) which display, respectively, the off- and on-diagonal projections of the jet-like three-particle correlation signals obtained with different levels of background subtraction. The increasing flow effect with increasing $\delta B_1$ pulls the off-diagonal peaks from the input cone angle $\pm1$ toward the flow peaks at $\pm\pi/4$, while an under-subtraction of background ($\delta B_1<0$) pushes the off-diagonal peaks away towards $\pm\pi/2$. The on-diagonal projection, on the other hand, only changes its magnitude but not its shape because the flow effect is unifrom along the on-diagonal direction as indicated in Figs.~\ref{fig6}(a) and (d).

Experimentally, the extracted conical emission angle is found not to be strongly affected by uncertainties in the background normalization level~\cite{Star_3part}. This suggests that the jet-correlation signal has variations with reaction plane similar to the medium background.

The complex structure of the lab-frame three-particle cumulant shown in Fig.~\ref{fig5}(c) is due to the irreducible flow correlation terms and the jet-flow cross-terms.
%The jet-like correlation part of the three-particle cumulant (but with over-subtraction of the background) is already shown in Fig.~\ref{fig3}(a).
The irreducible flow correlation terms, Eq.~(\ref{eq26}), is already shown in Fig.~\ref{fig4}(b). The contribution of the cross-terms between jet-correlation and anisotropic flow is shown in Fig.~\ref{fig8}(a). The three-particle cumulant excluding the jet-flow cross-terms is shown in Fig.~\ref{fig8}(b). Likewise, the three-particle cumulant excluding the irreducible flow correlation terms is shown in Fig.~\ref{fig8}(c). The three-particle cumulant excluding both the irreducible flow terms and the jet-flow cross-terms is show in Fig.~\ref{fig8}(d). The corresponding on- and off-diagonal projections are shown below each two-dimensional plot in panels (e)-(h), respectively. Of course, what is shown in Figs.~\ref{fig8}(d) and (h) is now in the same spirit of the jet-correlation method, subtracting specific background terms. Even with that, the input Mach-cone signal is not observable. This is because the subtracted backgrounds obtained with the average raw signal multiplicity ($\rho_1$) are too large. In fact, the cumulant with pure flow correlation and jet-flow cross-terms removed is very similar to that shown in Fig.~\ref{fig2}(c) for the uniform background case.

%Figure~\ref{fig5} (upper-left panel) shows the raw two-particle correlation function in black. The flow modulated background is shown in the dashed black curve. The red line shows the average of the correlation signal. The blue curve shows the normalized background by ZYAM. However, we note we have used $B_1=\Btrue$, as the two-particle jet-correlation $\Jhat_2(\dphi)$ we will use is not renormalized by ZYA1 or ZYAM. The three-particle cumulant result will be the same if we use the two-particle jet-correlation with normalized background and the corresponding new background level.

%%%%%%%%%%%%%%%%%%%%%%%%%%%%%%%%%%%%%%%%%%%%%%%%%%%%%%%%%%%%%%%%%%%%%%
\section{Summary}

We have described two analysis methods for three-particle azimuthal correlations between a high transverse momentum trigger particle and two softer associated particles: the {\em jet-correlation method} and the {\em cumulant method}. We point out two ways to analyze the cumulant, the {\em lab-frame cumulant} using azimuthal angle in the laboratory frame, and the {\rm reaction-plane-frame cumulant} using azimuthal angle relative to reaction plane in nucleus-nucleus collisions. We focused on the comparison between the jet-correlation method and the lab-frame cumulant method. We studied their differences analytically for two cases: one with a uniform background in azimuth and the other with a background including anisotropic flow. The major conclusions from our study are as follows:
\begin{enumerate}
\item[(1)] Cumulant is a mathematical construction; different mathematical representations of the same physics origin yield different cumulants. Extreme care should be taken in making physics interpretations of cumulant results.
\item[(2)] The reaction-plane-frame cumulant method and the jet-correlation method are identical if the background in the latter is obtained from mixed-events without normalization scaling. Since the azimuthal angle relative to reaction plane is a natural variable to use, we advocate the reaction-plane-frame method for future angular cumulant analyses in nucleus-nucleus collisions.
\item[(3)] 
The jet-correlation method has the jet-model built-in and is designed to study jet-like correlations; its interpretation is straightforward. The lab-frame cumulant method does not have a particular correlation model built-in and studies any kind of correlations; its interpretation is, however, difficult and has to involve a physical correlation model.
\item[(4)] For pure anisotropic flow without jet correlation, the jet-correlation method yields zero jet signal. The lab-frame cumulant method results in irreducible flow correlation terms and residual flow correlation due to non-Poisson statistics.
\item[(5)] With both anisotropic flow and jet correlation present, the jet-correlation method can construct the jet-correlation signal. However, the signal is distorted if incorrect background level is subtracted, due to any mismatch in anisotropies of jet-correlated particles and medium background particles. The lab-frame cumulant method results in jet-flow cross terms, in addition to the irreducible flow-correlation terms in (4) and the jet-correlation signal. The resultant lab-frame cumulant is very complex.
%\item[(2)] The main difference between the two methods lies in the different background subtraction schemes.
\item[(6)] For narrow jet peaks, both methods can identify Mach-cone structures, although the magnitudes and shapes of the correlation functions are different. For broad and more or less flat two-particle jet-correlations on the away side, as measured in RHIC experiments, the jet-correlation method can still identify Mach-cone structures, while the cumulant method fails. 
\end{enumerate}

%%%%%%%%%%%%%%%%%%%%%%%%%%%%%%%%%%%%%%%%%%%%%%%%%%%%%%%%%%%%%%%%%%%%%%
\section*{Acknowledgments}

We thank our STAR collaborators, in particular, Dr.~Marco van Leeuwen, Dr.~Claude Pruneau, and Dr.~Sergei Voloshin for valuable discussions. We thank Dr.~Andrew Hirsch for a careful reading of the manuscript. This work is supported by U.S. Department of Energy under Grant Nos. DE-FG02-02ER41219 and DE-FG02-88ER40412.

%%%%%%%%%%%%%%%%%%%%%%%%%%%%%%%%%%%%%%%%%%%%%%%%%%%%%%%%%%%%%%%%%%%%%%

\end{document}